\documentclass[letterpaper,dvipsnames]{article}
\pdfoutput=1
\usepackage{jheppub}
\usepackage{amsmath}
\usepackage{graphicx}% Include figure files
\usepackage{tikz}
\usetikzlibrary{arrows.meta, positioning}
\usepackage{array}
\usepackage{xcolor}
\usepackage{ulem}
\usepackage{slashed}
\usepackage{afterpage}
\usepackage{appendix}
\usepackage{multirow}
\usepackage{float}
\usepackage{mathtools}
\usepackage{scalerel}

\mathtoolsset{centercolon}

\usepackage{multirow}

\newcommand{\GeV}{\ {\rm GeV}}

\newcommand{\KK}{{\rm KK}}

\newcommand{\aGUT}{\alpha_{\scaleto{\rm GUT}{4pt}}}
\newcommand{\GGUT}{\mathcal{G}}
\newcommand{\GUT}{{\mathbb{G}}}
\newcommand{\MGUT}{M_{\scaleto{\rm GUT}{4pt}}}

\title{Axion couplings in Orbifold GUTs}

\author[a, b]{Prateek Agrawal,}
\author[c]{Michael Nee}
\author[b, d]{and Mario Reig}

\affiliation[a]{Department of Physics, University of California, Santa Barbara, CA 93106, USA}
\affiliation[b]{Rudolf Peierls Centre for Theoretical Physics, 
University of Oxford, Parks Road, Oxford OX1 3PU, United Kingdom}
\affiliation[c]{Department of Physics, Harvard University, Cambridge, MA, 02138, USA}
\affiliation[d]{Theoretical Physics Department, CERN, 1211 Geneva 23, Switzerland}
\emailAdd{prateek.agrawal@physics.ox.ac.uk}
\emailAdd{mnee@fas.harvard.edu}
\emailAdd{mario.reig.lopez@cern.ch}

\abstract{We consider the coupling of axions to gauge bosons in higher-dimensional Grand Unified Theories (GUT) inspired by string theory constructions with D-branes on orbifold singularities, the so-called orbifold GUTs. Due to their topological properties, axion couplings to gauge bosons are independent of the gauge symmetry reduction mechanism and the background geometry -- they only depend on the embedding of the SM into the UV gauge group. 
There are two kinds of axions in this class of theories: axions coming from gauge fields in the bulk and axions localised on the boundaries. The axion-photon coupling for the bulk axions coincide with the results from 4-dimensional GUTs, where axions which couple to photons necessarily couple also to QCD. 
The brane-localised axions can couple to photons independently of the QCD coupling, but if gauge couplings are approximately
unified, they have a much more severe quality problem. We show that in generic UV completions they are expected to get large masses from instantons on the boundaries.
This means that the ratio $g_{a\gamma\gamma}/m_a$ is below (or equal to) the QCD axion value for all axions in orbifold GUTs, as is the case for unified theories in 4d.}

\begin{document}

\maketitle

\section{Introduction}

\label{sec:introduction}

The quantum numbers of the Standard Model (SM) particles appear somewhat ad-hoc from a low energy perspective, but are elegantly explained in Grand Unified Theories (GUTs)~\cite{Georgi:1974sy, Fritzsch:1974nn}. One particular puzzle is why electric charges of colour-neutral particles are quantised in units of the electron charge. This, and several other predictions -- such as the prediction of the relative strength of the gauge interactions at low energies, all follow naturally from the embedding of the SM gauge group into a unified gauge group in the UV.

Despite this success, GUTs also predict incorrect ratios for some fermion masses and present other  problems, such as the doublet-triplet splitting problem. Addressing these challenges motivates the study of orbifold GUTs~\cite{Kawamura:1999nj,Barbieri:2000vh,Kawamura:2000ev,Altarelli:2001qj,Hall:2001pg,Nomura:2001mf, Hebecker:2001wq,Hebecker:2001jb,Asaka:2001eh, Hall:2001xr, Dermisek:2001hp,Hall:2001xb, Hall:2002ci,Kim:2002im,Kim:2004vk, Alciati:2005ur}, extra dimensional models where the GUT gauge group is broken by the boundary conditions satisfied by the gauge fields. 
The simplest effective theory that captures the features of these theories is a 5-dimensional theory with a unified gauge group in the bulk, and two 4-dimensional boundaries. Degrees of freedom localised to the boundaries may only come in representations of the boundary gauge group, not necessarily the gauge group of the bulk.
This offers a simple resolution to the doublet-triplet splitting problem, for example, as a doublet Higgs can be introduced as a boundary degree of freedom without its problematic triplet partner~\cite{Hebecker:2001wq}. The Higgs triplet can also be removed from the spectrum by projecting out the zero mode~\cite{Kawamura:2000ev}. Other unwanted GUT predictions such as fermion mass relations can similarly be rectified~\cite{Hall:2001pg}.
In these models, the gauge couplings are only approximately unified in the UV, due to brane-localised terms for the SM gauge fields. If $g_5$ is the (unified) dimensionful gauge coupling in the bulk, and $g_{\rm{brane,\,}i}$ the brane-localised coupling of the $i$'th SM gauge group, the requirement for the couplings to come close to unifying in the UV is
\begin{align}
    g_{\rm{brane,\,}i}^2 \gg \frac{g_5^2}{l_5} \, ,
    \label{eq:apparent_unification_intro}
\end{align}
where $l_5$ is the size of the extra dimension.

In previous work we have shown that the axion couplings to gauge bosons are sensitive to unification both in 4d GUTs~\cite{Agrawal:2022lsp} and weakly coupled heterotic string theory~\cite{Agrawal:2024ejr} (see also \cite{Reig:2025dqb} for models with non-standard embedding). This is because the axion-gauge boson coupling is topological, so is unaffected by renormalisation group (RG) flow. Unification implies that there is only one such coupling of axions to the full GUT gauge group in the UV, and the insensitivity to RG flow implies it remains unchanged in the IR. As a result any axion, $a$, has a photon coupling $g_{a \gamma \gamma}$ and mass $m_a$ which satisfy 
\begin{equation}\label{eq:coupling-to-mass_ratio}
  \frac{g_{a \gamma \gamma}}{m_a} \lesssim \frac{\alpha_{\rm em}}{2\pi}\left(\frac{E}{N}-1.92 \right) \frac{1}{m_\pi f_\pi} \, .
\end{equation}
Many ongoing and future experiments~\cite{Marsh:2018dlj,Lawson:2019brd,Beurthey:2020yuq,Schutte-Engel:2021bqm,DMRadio:2022pkf,Aja:2022csb,Bourhill:2022alm,ALPHA:2022rxj,Oshima:2023csb,DeMiguel:2023nmz,Ahyoune:2023gfw,Alesini:2023qed,BREAD:2023xhc,CAST:2024eil,Kalia:2024eml,Friel:2024shg}, including cosmic birefringence measurements~\cite{Minami:2020odp,Eskilt:2022cff,Diego-Palazuelos:2023mpy,Diego-Palazuelos:2025dmh}, are looking for light axions coupled to photons.  Any signal in these experiments would be incompatible with the prediction from unified theories.

The bound in eq.~\eqref{eq:coupling-to-mass_ratio} comes from the fact that a single axion couples to all gauge bosons, including gluons, via the anomaly, with other axions obtaining a coupling from mixing. The same ratio $E/N$ then sets the ratio of the gluon coupling to photon coupling for axions which are GUT singlets. In a GUT, if an axion couples to photons but does not couple to gluons, it corresponds to a symmetry generator that does not commute with the GUT gauge group. This means it gets a perturbative mass which exceeds the typical QCD axion masses by many orders of magnitude~\cite{Agrawal:2022lsp}. GUT-\textit{charged} axions thus satisfy~\eqref{eq:coupling-to-mass_ratio} due to their large perturbative mass. Equation~\eqref{eq:coupling-to-mass_ratio} therefore describes a robust bound in the axion parameter space satisfied by unified theories. The precise bound depends on the ratio of anomaly coefficients $E/N$; for the standard embedding of the SM into a simple gauge group $E/N=8/3$.\footnote{For a simple group the ratio of anomaly coefficients depends on the indices of embedding as $\frac{E}{N}=\frac{k_1+k_2}{k_3}$~\cite{Agrawal:2022lsp}. In 4d it can be shown that avoiding chiral exotica leads to $k_2=k_3=$ 1 or 3 but no unified theory with $k_2=k_3=3$ is known. The hypercharge normalisation is fixed to be $k_1=5/3$ to give the observed values of electric charges, which then gives the standard value $\frac{E}{N}=\frac{8}{3}$. More possibilities appear in higher-dimensional theories and string theory~\cite{Font:1990uw,Dienes:1996yh}. } 

In this paper we study the axion couplings to gauge bosons in the context of orbifold GUTs. There are two distinct cases. In orbifold compactifications without D-branes on the orbifold singularity, the axion coupling to the SM gauge bosons is GUT symmetric, and breaking of the gauge symmetry by the orbifold boundary conditions does not change the prediction in equation~\eqref{eq:coupling-to-mass_ratio}. In the case where there are D-branes localized on the singularity, there are two qualitatively different kinds of axions -- higher-form axions coming from gauge fields that propagate in the bulk and axions localised on the boundaries. Boundary-localised axions can couple to photons without coupling to gluons, whereas the bulk higher-form axions couple GUT symmetrically.

Axions on the boundary obtain a mass from brane-localised UV instantons. In situations where gauge couplings approximately unify, the brane-localised gauge couplings must satisfy equation~\eqref{eq:apparent_unification_intro}, implying that the brane-localised instantons are unsuppressed due to $g_{\rm{brane, }\,i}$ being relatively large. Importantly, these instantons are expected to be present even for $U(1)$ gauge groups, as is known in string theory~\cite{Blumenhagen:2009qh,Polchinski:1998rr,Ibanez:2012zz} and more recently has been shown to occur also in field theory~\cite{Fan:2021ntg,GarciaGarcia:2025uub}.
This results in a mass for boundary-localised axions which is much larger than the expected mass for the QCD axion, and in many cases lies close to the Kaluza-Klein (KK) scale, $M_{\rm KK}\sim l_5^{-1}$. Although brane-localised axions can couple independently to QED and QCD, they will also satisfy the bound~\eqref{eq:coupling-to-mass_ratio} due to their large masses. Once these non-perturbative effects are taken into account, we find that in any orbifold GUT theory where gauge couplings approximately unify all axions satisfy eq.~\eqref{eq:coupling-to-mass_ratio}.

There is also the possibility that bulk and boundary axions mix, leading to a `higher axion' which is a linear combination of the two~\cite{Loladze:2025uvf,Petrossian-Byrne:2025mto}. The orthogonal linear combination is eaten by a $U(1)$ gauge boson, where the axion being eaten is determined by the decay constants of the axions which mix. In this case, anomaly cancellation demands that the coupling of the higher-axion is GUT symmetric and so satisfies the bound~\eqref{eq:coupling-to-mass_ratio}.

The class of orbifold GUT models that we consider resemble type-II string theory models using D-branes (see \cite{Uranga:933469} for a review). In these theories, the bulk and boundary axions come from the Ramond-Ramond $p-$form fields in the closed string spectrum, while the higher axion (or open string axion in the type-II theory) comes from charged chiral matter living at the intersection of D-branes~\cite{Cicoli:2013cha,Cicoli:2013ana,Allahverdi:2014ppa,Petrossian-Byrne:2025mto}\footnote{Similar situations may arise in heterotic string theory with pseudo-
$U(1)$s~\cite{Dine:1987xk}.}. The instantons in these constructions come  from Euclidean D$q$-branes wrapping compact $(q+1)$-cycles, known as D-instantons~\cite{Blumenhagen:2009qh}. These objects are fully localised in four dimensions and reduce to the (small) gauge theory instantons in the low energy theory. 

Our results show that in orbifold GUTs with a GUT-symmetric bulk and apparent unification, all axions coupled to photons are on top of or below the QCD line -- even if the 4d EFT is never unified. The same conclusion applies to similar string models including type-II compactifications where the GUT gauge group is broken on a boundary. In minimal theories of this type, the decay constant of bulk axions is linked to the size of the extra dimension~\cite{Choi:2003wr}, preferring a decay constant around $f_a \approx l_5^{-1} = M_{\rm GUT}$.\footnote{In supersymmetric GUTs, $M_{\rm GUT} \approx 10^{16}$ GeV, but lower decay constants may be allowed if $l_5^{-1}\ll 10^{16}$ GeV. This requires additional model building to avoid constraints from proton decay, usually by separating SM fermions into different GUT representations and localising them in different parts of the bulk~\cite{Arkani-Hamed:1999ylh,Aldazabal:2000cn,Gherghetta:2000qt}.} For this reason, the QCD axion in orbifold GUTs is a well-motivated target for different planned experiments with sensitivities to axion masses near the neV scale. These include the ABRACADABRA experiment~\cite{Kahn:2016aff,Ouellet:2018beu,Ouellet:2019tlz,Salemi:2021gck}, DMRadio~\cite{DMRadio:2022jfv,DMRadio:2022pkf,Benabou:2022qpv,DMRadio:2023igr}, superconducting cavities~\cite{Berlin:2019ahk,Giaccone:2022pke}, and the CASPER experiment~\cite{Graham:2013gfa,Budker:2013hfa,JacksonKimball:2017elr,Aybas:2021cdk,Dror:2022xpi}.

This paper is organised as follows. In section~\ref{sec:5D_EFT} we review orbifold GUTs with special attention to the conditions that allow to retain gauge coupling unification. In section~\ref{sec:axions_in_5d_orbifoldGUT} we describe axions in these theories. We describe axions from higher-form gauge fields that propagate in the GUT bulk as well as axions localised at the orbifold singularity. In addition, we also consider cases where there exist brane-localised fermions inducing chiral suppression to the axion potential, cases where $l_5^{-1}\ll M_{\rm GUT}$, and the case where the bulk and brane localised axion mix. In section~\ref{sec:dim_deconstruction_orbifold} we describe some of the features of the orbifold GUTs in terms of dimensional deconstruction. In section~\ref{sec:typeII} we mention a possible connection to type-IIB string models. We conclude in section~\ref{sec:conclusion}.

\section{Orbifold GUTs}
\label{sec:5D_EFT}
Theories with compact extra dimensions offer new ways to break gauge groups beyond the Higgs mechanism. This is familiar from string theory models where the UV gauge group can be broken to the SM by background profiles for gauge fields, Wilson lines or orbifold projections~\cite{Witten:1985xc,Green:1987mn}. A class of models which capture some of the features of these models are known as orbifold GUTs \cite{Kawamura:1999nj,Barbieri:2000vh,Kawamura:2000ev,Altarelli:2001qj,Hall:2001pg,Nomura:2001mf, Hebecker:2001wq,Hebecker:2001jb,Asaka:2001eh, Hall:2001xr, Dermisek:2001hp,Hall:2001xb, Hall:2002ci,Kim:2002im,Kim:2004vk, Alciati:2005ur}.

Orbifolds can be constructed from smooth manifolds by identifying points related by a discrete subgroup, $F$, of the isometries of the manifold. If $F$ does not act freely, which means there are fixed points under $F$, and these correspond to orbifold singularities. Boundary conditions on the orbifold singularities can be chosen so as to not commute with the internal symmetries of the theory on the manifold. This has the effect of breaking the symmetry by boundary conditions, which is equivalent to breaking the symmetry by discrete Wilson lines. 

Orbifold GUTs are characterised by a unified gauge group propagating in a compact extra dimension, which is broken to the SM by boundary conditions for the gauge fields. The appeal of this mechanism is that some of the problems associated with unified theories can be circumvented. The boundary conditions project out parts of the full GUT multiplet. The Higgs doublet can appear without the colour triplet partners that are projected out by the orbifold, thus resolving the doublet-triplet splitting problem~\cite{Kawamura:2000ev}. Proton decay is also suppressed in these models as baryon-violating operators do not appear at dimension~5~\cite{Altarelli:2001qj, Hall:2001pg}. The usually problematic Yukawa coupling relations between different matter sectors can be avoided.
All this can be achieved while preserving the appealing GUT explanations of the SM couplings and quantum numbers. 
We note that the success of the orbifold models described so far is reproduced by heterotic compactifications on smooth manifolds~\cite{Witten:1985xc}. 

Orbifold GUT models traditionally have one additional ingredient that introduces significant model building freedom. Intersecting D-brane models defined on orbifolds (or even on smooth manifolds) can have different degrees of freedom living in different parts of the compactification. In orbifold GUTs, D-branes can be placed on the orbifold singularity, translating to the freedom of adding gauge sectors on the boundary with arbitrary matter representations under the boundary gauge group. The low-energy gauge group in 4d includes the bulk gauge group left unbroken by the boundary conditions, and the brane gauge group, which is usually further broken down to the Standard Model. This is somewhat similar in spirit to the flipped $SU(5)$ model with a non-simple unification group~\cite{Barr:1981qv, Derendinger:1983aj}. The additional freedom loses some of the attractive features of the GUT. Gauge coupling unification requires that the brane-localised contributions are subleading to the unified gauge coupling in the bulk. This requirement we term `apparent unification', and corresponds to the condition~\eqref{eq:apparent_unification_intro}. The explanation of the chiral SM matter multiplets filling out simple GUT representations is also lost. The similarity with flipped $SU(5)$ implies that (as was the case for 4d flipped $SU(5)$ in~\cite{Agrawal:2022lsp}) the axion-photon coupling can in principle avoid the bound in equation~\eqref{eq:coupling-to-mass_ratio}. However, we show in this paper that demanding apparent unification also severely restricts this possibility.

The model we consider in this section is an $SU(5)$ GUT compactified on the $S^1 /( \mathbb{Z}_2 \times \mathbb{Z}_2')$ orbifold~\cite{Hebecker:2001wq}. We could consider more general gauge groups with appropriate orbifold projections, or even theories in dimensions higher than d=5,\footnote{See for example~\cite{Huang:2016dtj} for an example using a 6d rectangular orbifold.} but this simple example captures the relevant features of this class of models. One could also consider warped extra dimensions~\cite{Randall:1999ee, Randall:1999vf}, which again do not change the qualitative behaviour of the model provided the extra dimension has the topology of $S^1 /( \mathbb{Z}_2 \times \mathbb{Z}_2')$.

\subsection{Orbifold Construction}

The $S^1 /( \mathbb{Z}_2 \times \mathbb{Z}_2')$ orbifold can be constructed by beginning with a periodic co-ordinate $y$ and defining the action of the two $\mathbb{Z}_2$ factors on the co-ordinates as \cite{Hebecker:2001wq}
\begin{equation}
\begin{aligned}
        &\mathbb{Z}_2 : y  \to - y \,,\\
    &\mathbb{Z}_2' : y  \to - y - \pi R \, .    
\end{aligned}
\end{equation}
The symmetry group $\mathbb{Z}_2 \times \mathbb{Z}_2'$ does not act freely as the points $ y = 0, y = l_5 = \pi R/2$ are both fixed under the parity transformations. Because of these fixed points the resulting space is not a manifold but an orbifold. The parity transformations act on an arbitrary field $\phi$ as
\begin{equation}
\begin{aligned}
    \phi(x,-y -\pi R)&=P'_\phi\phi(x, y)\,,
    \\
    \phi(x, -y)&=P_\phi\phi(x,y)\,,
\end{aligned}
\end{equation}
where $P_\phi, P'_\phi$ are matrix representations of the two $\mathbb{Z}_2$ groups, i.e. $P_\phi^2={P'}_\phi^2=1$. 

Considering the $SU(5)$ case, the gauge fields transform as:
\begin{align}
    T^a A^a_M (x, y) =  P T^a P^{-1} A^a_M (x, -y) \, , 
\end{align}
and similarly for $P'$, where $T^a$ are $SU(5)$ generators. If we take the parity assignments for the gauge multiplet to be
\begin{equation}
\begin{aligned}
    \label{eq:orbifold_parities_z2xz2'}
    P &=\text{diag}(+1,+1,+1,+1,+1)\,,
    \\
    P' &=\text{diag}(-1,-1,-1,+1,+1)\,,
\end{aligned}
\end{equation}
then at $y=0$ the boundary conditions commute with $SU(5)$, while at $y = l_5$ the boundary conditions break it to the SM gauge group $SU(3) \times SU(2) \times U(1)$.

\subsection{Adding D-branes and approximate gauge coupling unification}
If we add branes at each fixed point, we get additional degrees of freedom -- additional gauge groups and charged matter -- localized at these points. 
One typical choice for the additional gauge group on the ``visible" brane (at $y = l_5$) is taken to be $SU(3)\times SU(2)\times U(1)$.
The kinetic terms of the gauge sector are 
\begin{equation}
    \mathcal{L}_{\rm gauge}=\int_0^{l_5} dy\left[-\frac{1}{2g_5^2}\text{Tr}[\GGUT^{MN}\GGUT_{MN}]-\frac{\delta(y)}{2g_0^2}\text{Tr}[\GGUT^{\mu\nu}\GGUT_{\mu\nu}]
    -\sum_i\frac{\delta(y-l_5)}{2g_{\rm{brane}, i}^2}\text{Tr}[F_{i}^{\mu\nu}F_{i\,\mu\nu}]\right]\,,\,\,i=1,2,3\,,
\label{eq:Leff}
\end{equation}
where $\GGUT$ refers to the field strength of the unified gauge group, while the $F_{i}$ refer to the field strengths of the brane-localised gauge groups.

We take the SM gauge group to be the diagonal combination of the unbroken bulk gauge group and the brane localized gauge group.
The effective 4d gauge couplings at the compactification scale, $l_5^{-1}$, are given by 
\begin{equation}
    \frac{1}{g_i^2}=\frac{1}{g_{ 0}^2}+\frac{l_5}{g_5^2}+\frac{1}{g_{\rm{brane}\,i}^2} \, .
    \label{eq:4dcouplings}
\end{equation}
This kind of matching condition is similar to theories with a non-standard embedding of the SM~\cite{Agrawal:2017ksf} or  dimensional deconstruction (see section \ref{sec:dim_deconstruction_orbifold} for a discussion in that context). As discussed in eq.~\eqref{eq:apparent_unification_intro} the gauge couplings approximately unify if the GUT-symmetric brane and bulk terms dominate the brane localised contribution
\begin{align}
   \frac{1}{g_{ 0}^2}+\frac{l_5}{g_5^2} \gg \frac{1}{g_{{\rm brane}\,i}^2} \, .
   \label{eq:apparentunifiction}
\end{align}
Situations which violate this relation lose the compelling prediction of the gauge couplings in the IR from unification.

At energies below the compactification scale, $E\lesssim l_5^{-1}$, gauge couplings run logarithmically, leading to the usual prediction of the gauge coupling value if we identify $l_5^{-1}$ with the GUT scale, $l_5^{-1}= M_{\rm KK}\sim \MGUT \approx 2\times 10^{16}$ GeV,
\begin{equation}
	\alpha_i(l_5^{-1})\sim \aGUT=1/25\,.
\end{equation} 
For the weak mixing angle we retain the standard GUT prediction $\sin^2\theta_w=3/8$ for the standard embedding of the SM into an $SU(5)$ (or similar) GUT. 

\section{Axions in Orbifold GUTs}\label{sec:axions_in_5d_orbifoldGUT}
We now focus on axions and their couplings to gauge bosons in orbifold GUTs. When there are no additional branes on the orbifold fixed points, the analysis is similar to that of heterotic axions~\cite{Agrawal:2024ejr,Reig:2025dqb} (see also~\cite{Leedom:2025mlr}). There is more to say when there are D-branes present. Both cases are analysed in detail in this section.

\subsection{Axions in orbifold GUTs without D-branes}
Without boundary degrees of freedom, the axion comes from the 5th component of a bulk gauge field~$A_M$, $a=\int dy A_5$. The axion obtains a coupling to the GUT field strength $\GGUT$ from a bulk Chern-Simons~(CS) term
\begin{equation}
	\label{eq:CS_5dim}
	S_{CS}^{(5)}=\frac{n}{16\pi^2}\int d^5x\, \epsilon^{MNPQR} A_M\text{tr}[\GGUT_{NP} \GGUT_{QR}]\,,
\end{equation}
where the CS coupling constant $n$ takes integer values. The action~\eqref{eq:CS_5dim} is independent of the background metric, meaning that the discussion is independent of possible warping of the extra dimension.

Since the axion couples  to the bulk GUT gauge group, after integrating over the extra dimension we find that the axion couples universally to the gauge bosons in the 4d EFT
\begin{align}
    \mathcal{L} &= \frac{na}{16\pi^2} \text{tr}[\GGUT \tilde \GGUT]= \frac{na}{16\pi^2} \sum_ik_i\text{tr}[G_i \tilde G_i] \, .
\end{align}
Here $G_i$ stands for the different field strengths of the bulk gauge bosons whose zero modes survive after taking into account the orbifold boundary conditions, while $k_i$ refers to the embedding level of the $i$'th SM gauge group into the GUT gauge group. Therefore our previous results for 4 dimensional GUTs~\cite{Agrawal:2022lsp} are naturally extended to axions coming from gauge fields with CS couplings of the type (\ref{eq:CS_5dim}). In particular, the unified gauge symmetry implies that $a$ gets a coupling to photons as well as a mass from QCD effects. Any other axion can inherit a coupling to photons by mixing with $a$ and the ratio of the photon coupling $g_{a \gamma \gamma}$ to the axion mass $m_a$ is bounded above by the QCD line, i.e. satisfies equation~\eqref{eq:coupling-to-mass_ratio}. As is standard in $SU(5)$ GUT theories, we have $k_2=k_3=1$ and $k_1=5/3$, leading to $E/N=8/3$.

The shift symmetry of $a$ is of exponentially good quality because it is inherited from the global $1$-form symmetry of $A_M$. For this reason, only non-local effects in the form of particles travelling around the extra dimension can break this global symmetry and any boundary-localised effects will not spoil the flatness of the axion potential. When charged particles with mass $M$ are heavier than the KK scale any shift symmetry breaking effect is exponentially suppressed~\cite{Arkani-Hamed:2003xts,Craig:2024dnl} 
\begin{equation}
    V_{\rm PQV}(a)\propto l_5^{-4}e^{-S}\cos (a+\delta_{\rm PQV})\,, \text{ with: } S=l_5 M\,.
\end{equation}
In the limit that $M\rightarrow M_s$, with $M_s$ being the UV cutoff of the higher-dimensional theory (which corresponds to the string scale in a string UV completion), then the instanton action approaches the UV gauge instanton action, $S=\frac{2\pi}{\aGUT}$, and the quality of the axion is good enough to solve the strong CP problem. 

\subsection{Brane-localised axions}\label{sec:brane_loc_ALPs}
The presence of branes on the orbifold singularities means that axions can also be fields localised to the SM brane, where the full GUT symmetry is not respected. The brane-localised axion, which we denote $b$, can have distinct couplings to each of the SM gauge groups, breaking the relation between the axion decay constant and the mass in equation~\eqref{eq:coupling-to-mass_ratio}. For example, there could be a coupling $bF\tilde F$, where $F$ is the field strength of the localised $U(1)$. In principle, this opens up the possibility of having a light axion coupled to photons without a mass generated by QCD. 

The requirement of apparent unification \eqref{eq:apparentunifiction} leads to the axion-like particle (ALP) having a large mass, however, due to UV instantons associated with the gauge sector on the brane. In UV completions of the orbifold theory these instantons are expected to be present even for a $U(1)$ gauge group. While the details of these $U(1)$ instantons requires additional assumptions about the UV completion, in section \ref{sec:typeII} we will see that such non-perturbative effects are present in $D$-brane models~\cite{Blumenhagen:2009qh,Polchinski:1998rr,Ibanez:2012zz} (which are natural string UV completions of orbifold GUTs) and similar instanton effects are expected in other UV completions. Furthermore, as argued more recently in~\cite{Fan:2021ntg,GarciaGarcia:2025uub}, magnetic monopole loops can also play the role of the small $U(1)$ gauge instantons that we consider.
The brane-localised UV instantons generate irreducible contributions to the axion potential of the form\footnote{In order to be conservative we assume that the instanton contribution is dominated by instanton sizes $\rho \sim l_5$, leading to an axion potential that scales as the fourth power of the KK scale. }
\begin{equation}
	\label{eq:V_inst}
	V_{\rm inst}(b) = K\,l_5^{-4}e^{-S_{\rm inst}}\cos(b/f_b)\,.
\end{equation}
Here $K$ accounts for possible chiral suppression from light brane fermions, whose zero modes can be lifted by mass insertions or via Yukawa interactions with scalars, and the instanton action is 
\begin{equation}
	S_{\rm inst}=\frac{8\pi^2}{g_{{\rm brane}\,i}^2 (l_5^{-1})} \, . 
	\label{eq:inst_act}
\end{equation}
We note that this formula is found to hold for non-perturbative effects in the $U(1)$ examples discussed in ref.'s~\cite{Fan:2021ntg,GarciaGarcia:2025uub} as well as for instanton contributions from a non-abelian gauge theory.

We find that the requirement of apparent unification in orbifold GUTs (see~\eqref{eq:apparentunifiction}) implies that the instantons which generate the axion potential are unsuppressed. Using eq.~\eqref{eq:V_inst} we can derive the coupling-to-mass ratio in a way which is independent of $f_b$  
\begin{equation}\label{eq:coupling_to_mass_loc_ALP}
    \frac{g_{b\gamma\gamma}}{m_b}= \frac{\alpha_{\rm em}}{2\pi}\frac{1}{\sqrt{K}}\frac{1}{l_5^{-2}e^{-S_{\rm inst}/2}} \, .
\end{equation}
To compare $g_{b\gamma\gamma}/m_b$ to the same quantity for the QCD axion, $g_{a \gamma \gamma}/m_a$, we define $r_b$ as the dimensionless ratio
\begin{align} 
    r_b \equiv \frac{g_{b\gamma\gamma}}{m_b} \times \left( \frac{g_{a \gamma \gamma}}{m_a} \right)_{\rm QCD}^{-1} \simeq 20\frac{g_{b\gamma\gamma}}{m_b} \text{ GeV}^{2} \, .
    \label{eq:rb_ratio}
\end{align}
The generic expectation is that when the instanton action is small this leads to axions with $g_{b\gamma\gamma}/m_b$ orders of magnitude below the QCD axion line, or $r_b \ll1$. To illustrate this with an explicit example, let us take a brane-localised ALP coupled only to $U(1)_Y$ with brane-localised coupling taken to be $\alpha_{1}^{-1} =3$. In this example gauge coupling unification is no worse than in the SM alone ($\sim 10 \%$ accuracy). Then the instanton contribution to the ALP mass is
\begin{align}
	m_b \simeq 8\times 10^{11} \GeV \times          \sqrt{K} 
	\left(\frac{l_5^{-1}}{10^{16} \GeV}\right)^2 
	\left(\frac{10^{16}  \GeV}{f_b}\right)  \, ,
	\label{eq:ALPmass_estimate}
\end{align}
leading to a ratio of photon coupling to mass of
\begin{align}
    \frac{g_{b\gamma\gamma}}{m_b} \simeq \frac{1.5}{\sqrt{K}} \times 10^{-31} \text{ GeV}^{-2}  \,  , \qquad
    \implies r_b \simeq \frac{3}{\sqrt{K}} \times 10^{-30} \, .
\end{align}
We see that, for brane-localised axions, the orbifold GUT predicts additional ALPs with $\frac{g_{b\gamma\gamma}}{m_b}\ll \frac{g_{a\gamma \gamma}}{m_a}$.

The brane-localised heavy ALPs are well beyond the reach of laboratory axion searches but might have an impact on cosmology if they decay into photons~\cite{Gendler:2023kjt,Yin:2025amn}. We also show our results together with experimental constraints in figure~\ref{fig:ParameterSpace}. If the heavy ALP also couples to gluons it can be efficiently produced in supernovae via the gluon coupling and subsequently detected through its decay into photons~\cite{Benabou:2024jlj}, see red dotted line in figure~\ref{fig:ParameterSpace}. If the coupling to gluons is sufficiently large they could also be searched for at colliders~\cite{Jaeckel:2015jla,Bauer:2017ris,Bedi:2025hbz}. 

For the remainder of this section we consider the possibilities of enhancing this ratio with significant chiral suppression (small $K$) and/or a small KK scale (large $l_5$). We find that minimal orbifold GUTs allow for scenarios where a higher-form axion behaves as a QCD axion and lies on top of the usual QCD axion line, while also providing heavy ALPs coupled to photons to the right of this line, as highlighted in table~\ref{tab:results}. Similar situations arise in orbifold theories where the SM is unified with a confining dark sector~\cite{Foster:2022ajl} or in 4d constructions with multiple axions and sizeable mixing~\cite{Gavela:2023tzu,Dunsky:2025sgz}. 

\subsubsection{Brane-localised fermions and chiral suppression}
\label{sec:chiral}
In this section we look at the possible chiral suppression of the axion potential~\eqref{eq:V_inst} by fermions localised on the SM brane. Brane-localised fermions $\Psi$ will have zero modes on the instanton background that must be saturated with mass insertions~\cite{tHooft:1976snw}.\footnote{If there is a scalar field with a Yukawa interaction of the form $y_\Psi\phi \bar \Psi \Psi$ it is possible to saturate the zero modes with scalar loops. In this case, $K\sim \frac{y_\Psi}{4\pi}$. In order to obtain a conservative estimate, we will assume that zero modes are saturated only by fermion mass insertions.} For $n_F$ fermions, each with mass $M_\Psi$ this leads to a suppression factor
\begin{align}
	K=(l_5M_\Psi)^{n_F}\, .
\end{align}

These fermions modify the running of the gauge couplings above $M_\Psi$. Being brane-localised, they do not need to come in complete GUT multiplets, potentially spoiling apparent unification. The contribution of these fermions to the evolution of the 4D effective coupling at one loop between $M_\Psi$ and $l_5^{-1}$ is
\begin{align}
	&\alpha^{-1}_i(l_5^{-1})=
	\alpha^{-1}_i(M_\Psi)
	-\frac{\beta_i^{(1)}}{2\pi}
	\log\left( l_5 M_\Psi  \right) \, ,
\end{align}
where the index $i$ labels the three SM gauge groups. $\beta^{(1)}$ the one-loop coefficient of the beta function, computed as usual
\begin{equation}
	\beta_i^{(1)}=\frac{11}{3}C_2(G_i)-\frac{2}{3}\sum_f T_f-\frac{1}{3}\sum_s T_s\,.
\end{equation}
Above $M_\Psi$ we can write the one-loop beta function as $\beta^{(1)}_i = \beta_{i,\,\rm{SM}} + \Delta \beta_i$, where $\beta_{i,\,\rm{SM}}$ is the contribution from SM (or MSSM) degrees of freedom and $\Delta \beta_i$ corresponds to the new, localised fermions. This allows to write the gauge coupling at the compactification scale in terms of the SM (or MSSM) contribution and the new localised fermions, $\alpha^{-1}_i(l_5^{-1}) = \alpha^{-1}_{i,\,\rm{SM}}(l_5^{-1}) +\Delta\alpha^{-1}_i$. 

The contribution of brane-localised fermions to the running of each gauge group reads 
\begin{align}
	\Delta \alpha_i^{-1}
	= - \frac{2 n_F T_i}{3\pi}\log\left(l_5 M_\Psi  \right)
    \simeq 
    -\frac{2T_i}{3\pi} \log K
    \,,
\end{align}
where $T_i$ is the Dynkin index of the representation of the fermion $\Psi$ with respect to the $i$-th gauge group. Imposing that these contributions to the gauge coupling at the compactification scale do not spoil apparent unification \eqref{eq:apparentunifiction} leads to an estimate for $K$
\begin{equation}\label{eq:non_universal_corrections}
	\log K \sim  3\pi \aGUT^{-1} \frac{\Delta \alpha_i^{-1}}{\aGUT^{-1}}
    \sim
    240 \frac{\Delta \alpha_i^{-1}}{\aGUT^{-1}}
	\,.
\end{equation}
One can then use this criterion to bound the allowed chiral suppression. As an illustrative example, taking the localised fermion correction to be $\Delta \alpha_i \sim 1$, corresponding to a few \% correction to unification, corresponds to chiral suppression factor from brane-localised fermions given by
\begin{align}
	K=(l_5M_\Psi)^{n_F} = 8 \times 10^{-5} \,.
\end{align}
Models with larger chiral suppression will present non-universal contributions to the gauge coupling running that spoils gauge coupling unification by more than an $\mathcal{O}(1)$ amount. The axion mass depends on the square root of this factor, $m_a\propto\sqrt{K}$. A theory with $\Delta\alpha_i^{-1}\sim O(1)$ leads to a reduction by a factor of $\sqrt{K}\sim 10^{-2}$ in comparison to the naive estimate without taking into account chiral suppression. However, $K$ is exponentially sensitive to the tuning $\Delta \alpha^{-1}$ we allow, and taking $\Delta \alpha^{-1}\sim 3$ ($\sim 10$\% correction to unification) leads to $K=5\times 10^{-13}$.

If the brane-localised fermions come in complete GUT representations, unification of couplings in the UV will not be affected. However, $\aGUT^{-1}$ decreases with the addition of new fermions charged under the GUT gauge group (beyond SM matter). Adding $n_F$ Dirac fermions in the fundamental of SU(5), for example, will induce a change to $\aGUT^{-1}$ given by
\begin{align}
	\aGUT^{-1}
	\simeq 25 + \frac{n_F}{3\pi}\log\left(l_5 M_\Psi\right)\,.
\end{align}
This has two important implications. First, the brane localised contributions to the gauge coupling become more important (see discussion around~\eqref{eq:apparentunifiction}), i.e. $\aGUT/\alpha_i$ grows, so unification is made worse for fixed $\alpha_i^{-1}$. Second, a smaller $\aGUT^{-1}$ decreases the UV instanton action, which may modify the QCD axion potential, possibly reintroducing the strong CP problem. 

We also note that chiral suppression from complete GUT multiplets requires that the different components of the GUT representation have the same mass. For the case of a fermion transforming as the fundamental of $SU(5)$, the triplet $\Psi_3\sim (3,1,1/3)$ and the doublet $\Psi_2\sim(1,2,-1/2)$ universal chiral suppression requires $M_3 = M_2$. If there is a hierarchy between $M_3$ and $M_2$ we return to the  previous, non-universal case (see discussion around \eqref{eq:non_universal_corrections}). For these reasons chiral suppression of the potential is insufficient to make the localised axion light if we assume that gauge couplings approximately unify near the compactification scale, as shown in figure~\eqref{fig:ParameterSpace}.

\subsubsection{Low compactification scale}

\label{sec:low_compact_scale}
If the compactification scale $l_5^{-1}$ is much lower than the previously assumed value $l_5^{-1}\sim \MGUT \sim 10^{16}$ GeV, then a brane localised ALP will obtain a lower mass from localised UV instantons according to~\eqref{eq:ALPmass_estimate}. However, the scale $l_5^{-1}$ cannot be arbitrarily small. If SM matter is allowed to propagate in the bulk, KK modes of the $X$, $Y$ GUT gauge bosons may induce proton decay unless $l_5^{-1}$ lies close to $\MGUT$. 
When SM fermions are brane localised their interaction with $X$, $Y$ KK modes vanish, but can couple through higher dimensional operators to $\partial_y X_\mu, \partial_y Y_\mu$. These interactions can lead to baryon violating interactions arising at dimension 6, with suppression by the scale $l_5^{-1}$~\cite{Hebecker:2001wq}. If the couplings to $\partial_y X_\mu, \partial_y Y_\mu$ are sufficiently small then constraints from proton decay can be avoided for lower $l_5^{-1}$.

\begin{figure}[t]
    \centering
    \includegraphics[scale=0.4]{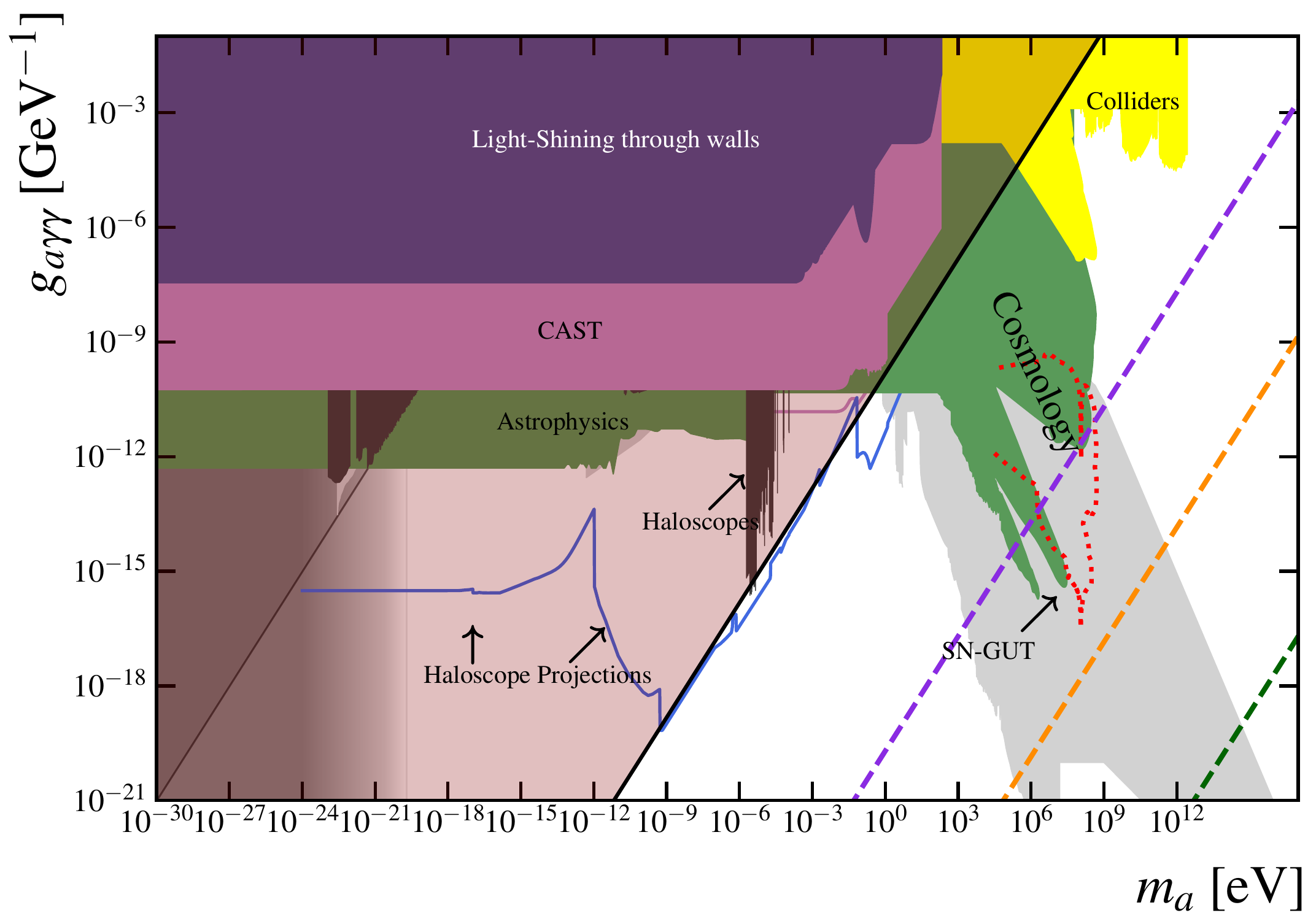}
    \caption{Preditions for $g_{a\gamma\gamma}$ and $m_a$ for the QCD axion (solid black) as well as brane-localised ALPs (dashed lines) with different choices of the relevant parameters according to the benchmarks in Table~\ref{tab:results}. \textcolor{Plum}{Purple} corresponds to low KK scale $l_5^{-1}\sim 10^9$ GeV and large chiral suppression, $K=5\times 10^{-13}$. \textcolor{orange}{Orange} corresponds to low KK scale and no chiral suppression, $K=1$. \textcolor{OliveGreen}{Dark green }corresponds to a model with large KK scale, $l_5^{-1}\sim 10^{16}$ GeV and $K=5\times 10^{-13}$. All cases assume $\Delta\alpha^{-1}=3$.
    The \textcolor{BrickRed}{dark red} region, corresponding to the parameter space where~\eqref{eq:coupling-to-mass_ratio} is violated, is not accessible to orbifold GUTs with apparent unification.  Experimental bounds adapted from~\cite{AxionLimits}.
   }
    \label{fig:ParameterSpace}
\end{figure}

Even if proton decay can be sufficiently suppressed, $l_5^{-1}$ still cannot be made too small without running into constraints from axion physics. The decay constant of a bulk axion is given by the KK scale, $f_a\sim l_5^{-1}$. Avoiding current constraints for axions coupled to photons one needs (see \cite{Caputo:2024oqc} for a recent review)
\begin{equation}
	f_a>10^9\text{ GeV}\,.
	\label{eq:fa_bound}
\end{equation}
Using eq.~\eqref{eq:coupling_to_mass_loc_ALP} we can easily estimate  the coupling-to-mass ratio for a given value of the compactification scale and the localised gauge coupling. 
Taking for example $l_5^{-1}\sim 10^9$ GeV, $\alpha^{-1}_{loc}=3$ (corresponding to O(10)\% precision for unification) and chiral suppression $K\sim 5\times 10^{-13}$ leads to the estimate $\frac{g_{b\gamma\gamma}}{m_b}=2\times 10^{-11} \text{ GeV}^{-2} $, see table~\ref{tab:results}.
This corresponds to an axion line to the right of the QCD axion which, for fixed coupling to photons, corresponds to axion masses several orders of magnitude heavier than the ones being proven at table top experiments such as haloscopes.
Figure~\ref{fig:ParameterSpace} shows the prediction for the axion coupling and mass for other different choices of the parameters. 

\begin{table}[t!]
\begin{center}
\begin{tabular}{c||c|c|c}
    % \hline
      \textbf{Scenario} &  $l_5^{-1}$ & $K$&  $r_b$   \\  
      % $\frac{g_{b\gamma\gamma}}{m_b} \times$ (GeV)$^2$   \\  
     \hline 
     Generic Expectation  &  $10^{16}$ GeV  &  $1$     &   $3\times 10^{-30}$  \\
     % $1.5\times 10^{-31}$  \\
     Low KK scale         &  $10^{9}$ GeV  &  $1$   &    $3\times 10^{-16}$  \\
     % $1.5\times 10^{-17}$  \\
     $K$ Suppression      &  $10^{16}$ GeV  &  $5 \times 10^{-13}$   &    $4\times 10^{-24}$  \\
     % $2\times 10^{-25}$  \\
     $K$ Suppression \& Low KK scale  &  $10^{9}$ GeV  &  $5 \times 10^{-13}$     &  $4 \times 10^{-10}$
     % $2 \times 10^{-11}$
     % \\      \hline 
\end{tabular}
\end{center}
\caption{Values for the ratio of photon-coupling to mass for the brane-localised axion in different benchmark scenarios discussed in the text, the expression is given in eq.~\eqref{eq:coupling_to_mass_loc_ALP}. The generic expectation assumes no chiral suppression and KK scale ($l_5$) at the GUT scale. We also consider lowering the KK scale, adding chiral suppression and doing both at the same time. The final column shows the ratio $r_b$, defined in eq.~\eqref{eq:rb_ratio}, highlighting that brane-localised ALPs have exponentially small couplings relative to the QCD axion expectation for each of the cases we consider.
% Comparing to the QCD axion value $g_{a\gamma \gamma}/m_a \simeq 0.05 \text{ GeV}^{-2}$, each of the cases leads to an axion with exponentially small coupling relative to a QCD axion of the same mass.
In each case we allow for $\Delta \alpha_i^{-1} = 3$, corresponding to corrections of around $10 \%$ to the unified value. }
\label{tab:results}
\end{table}

\subsection{Mixing between higher-form and brane-localized axions: higher axions}\label{sec:higher-axion}
It is also possible to have scenarios where the higher-form axion from $A_M$ and a field theory axion localised on the boundary mix. This may occur in string constructions where there are pseudo-anomalous $U(1)$ gauge symmetries~\cite{Choi:2011xt,Choi:2014uaa,Buchbinder:2014qca,Loladze:2025uvf} with the anomalies cancelled via a generalised Green-Schwarz mechanism~\cite{Green:1984sg}. The anomaly cancellation mechanism requires a non-trivial cancellation between a boundary localised anomaly and a contribution from a bulk CS interaction. This results in a perturbatively massless axion, $\theta$, -- which we call the higher-axion -- with contributions from $A_5$ as well as from brane-localised axions.  

In the case that the bulk is GUT symmetric but the boundary is not, it is interesting to ask about the couplings of the higher-axion as well as the non-perturbative effects breaking its shift-symmetry. In addition to the bulk gauge $U(1)$ where the higher-form axion originates from, the construction we consider requires an additional $U(1)_C$ boundary gauge symmetry under which a localised complex scalar, $\Phi=|\Phi|e^{ib}$, is charged (see~\cite{Petrossian-Byrne:2025mto} for a related construction). The gauge symmetry $U(1)_C$ forbids local operators breaking the shift symmetry of $b$. 

The mechanism requires that the boundary $U(1)_C$ has anomalies which are canceled by shifting the bulk axion
\begin{equation}\label{eq:fancy_gauge_transf}
    A_5 \rightarrow A_5 + \delta(y)\lambda_C\,.
\end{equation}
This transformation links both the bulk and boundary $U(1)$ gauge symmetries and has two important implications. First, the bulk CS interaction \eqref{eq:CS_5dim} can cancel localised $U(1)_C$ anomalies induced by chiral fermions $\psi_i$ on the boundary. The second is that the anomalous gauge symmetry will be spontaneously broken, and the associated gauge boson $C_\mu$ will get a mass by eating a linear combination of the higher-form axion $\int dy A_5$ and the phase $b$. The exact linear combination which is eaten depends on the hierarchy between the decay constants, which are set by the scalar vev ($|\Phi|$) and the KK scale. 

The orthogonal (un-eaten) combination is the higher axion, $\theta$. This is parametrised by the vacuum manifold~\cite{Loladze:2025uvf}
\begin{equation}
    U(1)_{\rm PQ}=\frac{U(1)^{(0)}\times U(1)^{{\color{orange}(1\downarrow0)}}}{U(1)_{\rm gauge}}\,,
\end{equation}
which reflects the fact that the shift symmetry of $\theta$ originates from the mixing between the global higher-form symmetry, $U(1)^{{\color{orange}(1\downarrow0)}}$, as well as the boundary localised symmetry, $U(1)^{(0)}$. One can see that the boundary instantons do not generate a mass for $\theta$ as they preserve the more robust 1-form symmetry. Any contributions to the mass will be exponentially suppressed by the size of the extra dimension.

Let us now consider the couplings to gauge bosons of the higher axion, $\theta$. The key ingredient is the mixing between the higher-form axion and the localised axion, which is induced by the gauge transformation~\eqref{eq:fancy_gauge_transf}. This indicates that the shift of the  gauge boson $A_M$ can cancel (brane localised) GUT universal anomalies (of the form $[\GUT_{\rm GUT}]^2\times U(1)_C$) via the CS term in eq.~\eqref{eq:CS_5dim}. In turn this implies that the coupling of $\theta$ to gauge bosons will be GUT-symmetric. This is true even if there is not a unified effective theory below the scale $l_5^{-1}$ and even if the un-eaten linear combination (higher-axion) is mainly given by the phase $b$. The higher axion couplings are therefore given by
\begin{equation}
    %\frac{a}{f_a} 
  \mathcal{L}_\theta= \theta  \left (\frac{k_1}{16\pi^2} B\tilde B+\frac{k_2}{16\pi^2} \text{tr}[ W\tilde W]+\frac{k_3}{16\pi^2}\text{tr}[G\tilde G ]\right )\,.
\end{equation}

In the limit $l_5^{-1}\gg |\Phi|$, the higher-axion decay constant is dominated by the complex scalar and $f_\theta\sim |\Phi|$. This is in close analogy to the axion couplings of field theoretic axions in heterotic strings studied in~\cite{Agrawal:2024ejr, Reig:2025dqb,Buchbinder:2014qca,Choi:2011xt,Choi:2014uaa}. The low-energy phenomenology of $\theta$ is similar to the higher-form axion described above, with the difference that it may lead to a post-inflationary axion cosmology.\footnote{See~\cite{Benabou:2023npn} for a discussion on the difficulty to form axion strings for axions coming from higher-form gauge fields.} In that case, recent simulations point to a QCD axion decay constant around $f_\theta\sim 10^{10}-10^{11}$ GeV, offering relatively accurate predictions for the dark matter QCD axion mass~\cite{Saikawa:2024bta,Correia:2024cpk,Benabou:2024msj}.

\section{Orbifold deconstruction and localised axions}
\label{sec:dim_deconstruction_orbifold}
\begin{figure}
    \centering
\begin{tikzpicture}[
  node distance=3cm,
  every node/.style={font=\small},
  gauge/.style={circle, draw, thick, minimum size=1.2cm, inner sep=0pt, align=center},
  >=Stealth
]

% Gauge group nodes
\node[gauge, draw=red, text=red] (G1) {\( SU(5)_1 \)};
\node[gauge, right of=G1] (G2) {\( SU(5)_2 \)};
\node[, right of=G2] (dots) {\( \cdots \)};
\node[gauge, right of=dots] (G4) {\( SU(5)_{N} \)};
\node[gauge, draw=red, text=red, right of=G4] (G5) {\( 321 \)};

% Bifundamental scalar links
\draw[->, thick] (G1) -- node[above] {\( \Sigma_{1} \)} (G2);
\draw[->, thick] (G2) -- node[above] {\( \Sigma_{2} \)} (dots);
\draw[->, thick] (G2) -- (dots);
\draw[->, thick] (dots) -- node[above] {\( \Sigma_{N-1} \)} (G4);
\draw[->, thick] (G4) -- node[above] {\( \Sigma_{N} \)} (G5);

\end{tikzpicture}
    \caption{\textbf{Deconstructed orbifold GUT.} Each site has a $SU(5)$ gauge sector which is Higgsed down to the diagonal subgroup via the vev of bi-fundamental (link) fields, $\Sigma_i$. The last boundary site, in red, contains a $SU(3)\times SU(2)\times U(1)$ gauge sector which is physically relevant (not an artifact from the deconstruction picture). It may contain fields transforming in SM representations that do not fill out a full $SU(5)$ multiplet. This site can also contain axion-like particles which couple to gauge bosons in a non GUT-symmetric way. As explained in the text these ALPs get a large mass due to non-perturbative effects with a small instanton action. The action can be calculated by using the UV gauge coupling $S_{i}\sim \frac{2\pi}{\alpha_i}$, with $\alpha_i$ the localised gauge coupling. }
\label{fig:deconstructed_orbifold}
\end{figure}
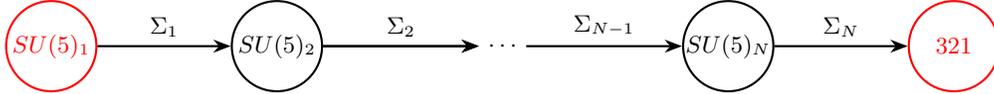

An understanding of the symmetry breaking patterns in orbifold GUTs can be arrived at by using dimensional deconstruction~\cite{Hill:2000mu,Arkani-Hamed:2001kyx}. This involves latticizing the 5th dimension and writing the theory as $N$ copies of a 4D gauge theory with next-to-neighbour interactions. There are link Higgs fields that transform as bifundamentals and break the symmetry on each pair of sites to the diagonal subgroup. Of course, 5D gravity is not captured in this formalism. The continuum theory is recovered at distances larger than the lattice spacing. To generalise to the orbifold case, we consider the case where the GUT bulk is deconstructed, and add two sites to the lattice which contain the gauge groups localised to each brane. This approach to orbifold theories was  considered in \cite{Csaki:2001qm}, and a cartoon is given in figure~\ref{fig:deconstructed_orbifold}.

The $SU(5)$ sites are connected by bi-fundamental fields $\Sigma_i$ that trigger the spontaneous breaking down to the diagonal subgroup, $\prod^N_iSU(5)_i\rightarrow SU(5)_{\rm diag}$, at the SSB scale, $\Lambda_{\rm SSB}$. The localised gauge sector has the same structure as the SM, $SU(3)\times SU(2)\times U(1)$. The resulting group %
\begin{equation}
SU(5)_{\rm diag}\times SU(3)_{\rm brane}\times SU(2)_{\rm brane}\times U(1)_{\rm brane}\,,
\end{equation}  
is broken down to the diagonal gauge group, $SU(3)_C\times SU(2)_L\times U(1)_Y$. 
The tree-level matching condition at the compactification scale determines the low-energy 4d gauge couplings

\begin{equation}\label{eq:matching}
\frac{1}{g_i^2}=\frac{l_5}{g_5^2}+\frac{1}{g_{\rm{brane},\, i}^2}\,.
\end{equation}
In the above equation,  $g_{\rm{brane},\, i}^2$ is the gauge coupling at the SM-like site (brane localized contribution) and $g_5$ is the 5D (dimensionful) gauge coupling of $SU(5)_{\rm diag}$. It is related to the gauge couplings at each lattice site, $g$, as $g_5^2=a g^2$, with $a$ the lattice spacing. Finally, the size of the extra dimension $l_5$ is related to the lattice spacing and the number of sites as $l_5=a N$. 

In order to recover apparent unification \eqref{eq:apparent_unification_intro}, one can simply assume that the volume-enhanced bulk coupling dominates the tree-level matching condition, $ l_5/g_5^2\gg g_{\rm{brane},\, i}^{-2}$. When this condition is satisfied we explain why the 4-dimensional SM gauge couplings approximately unify at the compactification scale, $\alpha_i^{4d}\approx \alpha_{\rm GUT}$, even though we never had a unified, single gauge group in the UV. Below the compactification scale, $E\lesssim l_5^{-1}$, the running of $g_i$ is logarithmic and we recover the weak mixing angle prediction.

The deconstructed picture can also clarify the origin of the UV instanton contributions to the potential. A 4d theory given by the product of $N$ $SU(5)$ gauge group factors would have instantons associated to each site, generated at the scale of symmetry breaking. Each of these instantons would have an action given by $S_i\sim 2\pi/ \alpha_i\sim 2\pi /(N\alpha_{\rm GUT})$. These instantons can reintroduce the strong-CP problem, even for $N=2$, if there are new CPV phases beyond those in the SM. However, the fractional instantons at each $SU(5)$ site are just artifacts of deconstruction. Only the product of $N$ fractional instantons -- one at each site -- has physical meaning~\cite{Poppitz:2002ac}.  The action is simply given by the sum of their actions 
\begin{equation}
S=\frac{2\pi N}{\alpha_i} =\frac{2\pi }{\alpha_{\rm GUT}}\,.
\end{equation}
We see that this corresponds to the action of the UV instanton  breaking the shift symmetry of a 5D bulk axion coupled to the full $SU(5)$ via the CS term (see equation~\eqref{eq:CS_5dim}). 

Importantly, the situation is qualitatively different at the boundary sites. This includes the last site of the deconstructed picture, where only the SM gauge symmetry is realised. Given that it corresponds to a defect stuck at the singular point of the higher-dimensional GUT bulk, there are physically relevant non-perturbative effects associated to it. We remark that these are not artifacts of deconstruction and result in shift-symmetry breaking effects that generate a potential for brane-localised axions. The instanton action is
\begin{equation}
S_i\sim \frac{2 \pi }{ \alpha_{i}}\,.
\end{equation}
with $\alpha_i=\frac{g_i^2}{4\pi}$ the localised gauge coupling at the SM-like site. As before, the instanton action is small, $S_i\sim O(1)$, if we require approximate unification of gauge couplings (see Eq. \ref{eq:matching}). Because of this, localised axions are expected to have a mass which can be as large as the KK scale when the gauge couplings approximately unify due to the dominant bulk contribution. Similar to the 5d model, this framework allows us to correlate precise unification of gauge couplings with large masses for brane localized axions -- that is, the only axions that can avoid the coupling to gluons, potentially violating the relation~\eqref{eq:coupling-to-mass_ratio}, get a large mass from UV instantons. 

We note, however, that in the field theoretic 4D deconstructed picture it seems that the axion coupled to $U(1)_{\rm brane}$ can remain massless due to the absence of $U(1)$ gauge instantons in 4d. However, as we saw in \ref{sec:brane_loc_ALPs}, in a UV complete theory with branes at orbifold singularities this will not be the case. Axions, understood as $0-$form gauge fields, will necessarily couple to instanton-like objects breaking their shift symmetry. In the case of $U(1)$ gauge sectors from a D3-brane, the instanton-like objects are D(-1)-branes~\cite{Blumenhagen:2009qh}, while in field theory they can be related to monopole loops~\cite{Fan:2021ntg,GarciaGarcia:2025uub}.

\section{Connection to string theory}
\label{sec:typeII}

The orbifold GUT models we consider in this work can be considered to be simplified models of compactifications of type-II string theory. Such constructions typically feature branes which support localised gauge theories, and use discrete symmetries (such as the orbifold projections we consider here) to break the bulk $\mathcal{N}=2$ supersymmetry down to $\mathcal{N}=1$. These models may also contain a large number of axions, depending on the complexity of the compactification manifold~\cite{Conlon:2006tq,Svrcek:2006yi,Arvanitaki:2009fg,Cicoli:2012sz,Hebecker:2018yxs,Demirtas:2018akl,Halverson:2019cmy,Mehta:2021pwf,Demirtas:2021gsq,Foster:2022ajl,Gendler:2023hwg,Gendler:2023kjt,Reece:2024wrn,Gendler:2024adn,Benabou:2025kgx, Cheng:2025ggf,Fallon:2025lvn}. 

\begin{figure}[t!]
\centering
\begin{tikzpicture}

\filldraw[fill=blue!50, draw=black, thick] (-1.2,0) -- (1.2,1.2) -- (1.2,6.0) -- (-1.2,4.8) -- cycle;

\filldraw[fill=blue!50, draw=black, thick] (7.8,0) -- (10.2,1.2) -- (10.2,6.0) -- (7.8,4.8) -- cycle;

\draw[dotted] (-1.2,0) -- (7.8,0);      % Bottom left corners
\draw[dotted] (1.2,1.2) -- (10.2,1.2);  % Bottom right corners
\draw[dotted] (1.2,6.0) -- (10.2,6.0);  % Top right corners
\draw[dotted] (-1.2,4.8) -- (7.8,4.8);      % Top left corners

\node at (0,3)  {$\Sigma_4^{\rm bdy 1} \cap\Sigma_4^{\rm bulk}$};  
\node at (9,3)  {$\Sigma_4^{\rm bdy 2} \cap\Sigma_4^{\rm bulk}$};  
\node at (4.5,3)  {$ \Sigma_4^{\rm bulk}$};  

\draw[->] (-2.5,-.5) -- (-2.5,.5) node[left] {$y_{3}$};  % y-axis
\draw[->] (-2.5,-.5) -- (-1.5,-.5) node[below]{ $y_{1, 2}$}; % x-axis
\draw[->] (-2.5,-.5) -- (-3,-1.1) node[left]{ $y_{4}$}; % z-axis (out of page)

\end{tikzpicture}
    \caption{Diagram showing the D-7 brane setup which gives an analogue of the orbifold GUT model. The stack of branes which form the bulk span the $y_{1 - 4}$ directions, while the boundary branes are extended in the $y_{3 -6}$ directions. The intersection of the branes is indicated by the shaded regions, which extend in the directions $y_{3, 4}$. The 5d theory is obtained by integrating over the co-ordinates $y_3 - y_6$ and one of $y_1, y_2$.}
    \label{fig:typeII}
\end{figure}
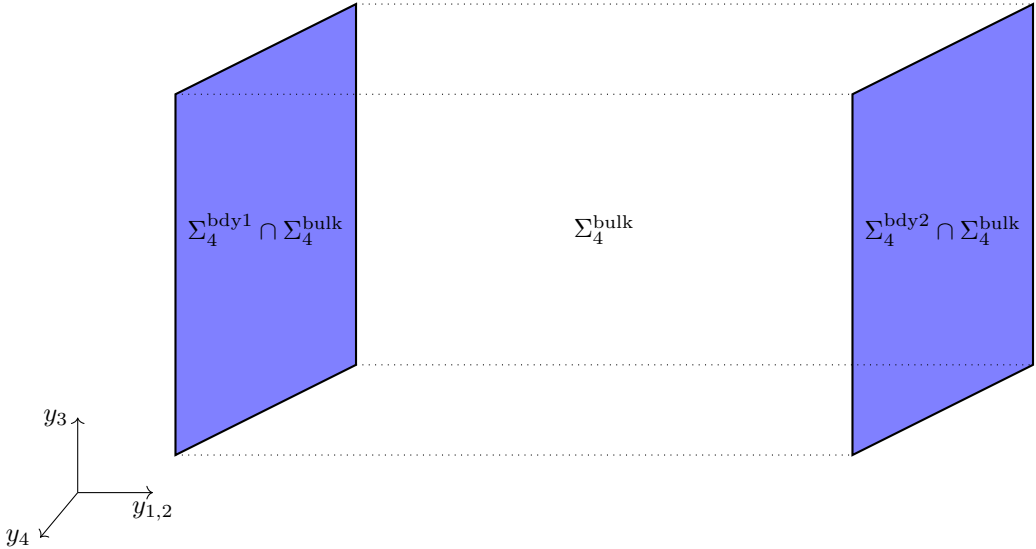

Here we discuss what is required for a type-II construction to lead to an EFT similar to the orbifold GUT we considered in the previous section. The setup we have in mind is shown in figure~\ref{fig:typeII}. The necessary ingredients are:
\begin{itemize}
    \item A set of branes on an internal cycle which host an $SU(5)$ gauge theory. The volume of this cycle, $\mathcal{V}_{\rm bulk}$, sets the bulk contribution to eq.~\eqref{eq:apparent_unification_intro}. 
    \item Two sets of branes which support the boundary theories. These live on cycles which intersect the bulk cycle. These branes allow for localised kinetic terms for the gauge fields and degrees of freedom which do not necessarily come in full multiplets of the bulk theory. The worldvolume of the branes also contain axions and non-perturbative effects breaking their 4d shift-symmetries.
    \item The volume, $\mathcal{V}_{\rm bulk}$, of the bulk cycle should be large compared to other cycles in the compactification manifold in string units. This is so that our 5d EFT, where we have kept just a single extra dimension, is valid. In such a UV completion, this requirement will also means that the bulk gauge coupling is weaker than the gauge couplings on the boundary branes, so apparent unification~\eqref{eq:apparent_unification_intro} is a natural expectation.
\end{itemize}

In the 5d orbifold GUT, we only have a single extra dimension and consider 3-branes extended only in the infinite dimensions on the boundaries. This could arise from a type-IIB compactification with D7-branes wrapped on a Calabi-Yau. 
As an example, D7-branes wrapped on a 4-cycle, $\Sigma_{\rm bdy}$, can host the boundary gauge theories. If the the volume, $\mathcal{V}_{\rm bdy}$, of $\Sigma_{\rm bdy}$ is small, then we can integrate over these 4 dimensions, leaving a six dimensional theory. An additional stack of D7-branes wrapped on a 4-cycle, $\Sigma_{\rm bulk}$, that overlaps the remaining two internal dimensions, $y_1$ and $y_2$ can support the bulk $SU(5)$ gauge theory. Due to the dimensionality of the cycles, $\Sigma_{\rm bulk}$ will overlap $\Sigma_{\rm bdy}$ along two dimensions, as shown in figure~\ref{fig:typeII}. If the geometry of the bulk is such that one direction ($y_1$) is larger than the other ($y_2$), we are left with an effective 5-dimensional theory after integrating over $y_2$. The validity of this EFT requires that the volume of the bulk spanned by $y_1, y_2$, $\mathcal{V}_{\rm bulk}$, is larger than $\mathcal{V}_{\rm bdy}$ in string units, and that $y_1\ll y_2$. As the inverse gauge couplings on the brane are proportional to $\mathcal{V}_{\rm bdy}$, apparent unification in fact requires that these volumes be small, close to $O(1)$ in string units.

Let us now discuss how GUT-symmetric axions, brane-localised ALPs $b^{(i)}$ and their corresponding shift symmetry breaking effects arise in this framework. The GUT symmetric axion arises from the RR four-form field integrated over the bulk cycle, $a=\int_{\Sigma_{\rm bulk}}C_4$, and apart from IR instantons it receives a shift-symmetry breaking potential from Euclidean D$3$-branes wrapping the bulk 4-cycle, $\Sigma_{\rm bulk}$ with an action scaling as $S\sim 2\pi/\alpha_{\rm GUT}$. On the other hand, non-universal ALPs arise from the RR four-form field integrated over the boundary cycle, $b^{(i)}=\int_{\Sigma_{\rm bdy}^{(i)}}C_4$. The coupling to gauge bosons of this ALP can be obtained from the CS part of the D$7$-brane action wrapping $\Sigma_{\rm bdy}^{(i)}$
\begin{equation}
    S_{CS}\subset \int_{M_4\times \Sigma_{\rm bdy}^{(i)}} C_4\wedge \sum _j n_j\text{tr}F_j^2 \,,
\end{equation}
where $j$ indicates the different gauge sectors from the different stacks of D$7$-branes wrapping $\Sigma_{\rm bdy}^{(i)}$, and $n_j$ is the integer that parametrizes the coupling to these gauge sectors (not necessarily GUT symmetric).
In this case, the ALP $b^{(i)}$ obtains shift-symmetry breaking effects from Euclidean D$3$-branes wrapping $\Sigma_{\rm bdy}^{(i)}$ with an action $S_{\rm ins}= 8\pi^2/g_{\rm brane}^2$. This coincides with the departure from the unified gauge coupling as discussed in section \ref{sec:brane_loc_ALPs}, see also Eq.~\eqref{eq:apparentunifiction}. As unification gets more precise, $S_{\rm inst}$ decreases and the ALP becomes heavier.

In this setup, both the bulk and boundary axions come from higher-form gauge fields -- the RR $C_4$ field in the case of theories with D7-branes -- integrated over the different cycles. This means that there is no difference in the source of the axions. However, the `higher axion' discussed above comes from a combination of a bulk axion and a boundary axion, which comes from an open string mode where the string ends on the bulk and boundary branes. The mixing occurs if there is a Green-Schwarz anomalous $U(1)$ in the EFT, which typically requires a $U(1)$ flux supported on the intersection of the branes. 

We also note that a setup such as ours does not occur in perturbative heterotic string models. In the weakly coupled heterotic theory there are no branes to host localised degrees of freedom. This means the gauge theory is the same everywhere in the internal dimensions. In the strongly-coupled (M-theory) limit, there are two gauge theories on the boundaries of the extra dimension but the bulk is purely gravitational~\cite{Horava:1995qa, Horava:1996ma}. 

\section{Discussion}\label{sec:conclusion}

In this work we determined the axion couplings to gauge bosons in orbifold GUTs. These theories famously solve long-standing issues of standard 4d unified theories such as the doublet-triplet splitting problem. We have shown that there exist two main sources of axions. The first is where the axion comes from a 1-form gauge field that propagates in the bulk. This axion couples to the GUT gauge bosons in the bulk via the CS term, so couples to both photons and gluons with relative strength $E/N$. As in standard GUTs this ratio is fixed by the embedding of the SM into the simple gauge group. Without branes at the orbifold fixed point only this bulk axion couples to gauge bosons with any additional axion receiving only mixing-suppressed couplings similar to 4d GUTs.

The presence of branes at the orbifold singularity introduces additional possibilities. In this case there can also be axions on the boundaries that couple independently to photons and to gluons.
These brane-localised ALPs, however, obtain a large mass from localised instantons, which necessarily have a small action due to the requirement that the couplings approximately unify. The mass of such axions increases as the unification of couplings near the KK scale gets more precise. This occurs because the contribution to the gauge coupling from the branes also controls the instanton action, which gets small as the couplings approximately unify. Our results are shown in table~\ref{tab:results}, where we compare the  $g_{a\gamma\gamma}/m_a$ ratio for brane-localised axions with a GUT symmetric QCD axion in different benchmark scenarios. In figure~\ref{fig:ParameterSpace} the predicted coupling-to-mass ratios, as well as the region that is not accessible in orbifold GUTs, are shown together with experimental constraints and prospects for the reach of future experiments.

In section~\ref{sec:typeII} we have discussed the possible realisation of the orbifold GUT models in type-II string theory. While we have not considered a complete stringy model, it is relatively straightforward to imagine a scenario with all the ingredients necessary to realise our orbifold GUT setup. Similar results for the axion parameter space would apply in these models.

In summary, orbifold GUTs contain axions coming from bulk gauge fields, which will couple like a QCD axion and lie on top of or below the usual QCD axion line. In minimal models where $M_{\KK}\sim M_{\rm GUT}$, the mass of the QCD axion is predicted to be around $m_a\approx 6\times 10^{-10}$ eV. There can also be additional, heavy ALPs coupled to photons to the right of this line, coming from axions on the boundary which receive large instanton contributions to the potential or from other bulk axions which couple via mixing with the QCD axion. These heavy ALPs are expected to have an interesting impact in astrophysics and cosmology. Just as in the usual 4d GUTs light axions coupled to photons breaking the relation \eqref{eq:coupling-to-mass_ratio} are also not present in orbifold GUTs, despite the presence of terms which explicitly violate the GUT symmetry. 

\section*{Acknowledgments}
% We thank Mario Reig for collaboration on the early and late stages of this project.
We thank Josh Benabou, Arthur Hebecker, Anson Hook, Junwu Huang, Andre Lukas, John March-Russell, and \textcolor{orange}{Arthur Platschorre} for useful conversations and discussions. 
PA is supported by the STFC under Grant No. ST/T000864/1.
MN is supported in part by NSF Award PHY-2310717, and would like to acknowledge GRASP Initiative funding provided by Harvard University.

\bibliographystyle{utphys}
\bibliography{newrefs_axion}

\providecommand{\href}[2]{#2}\begingroup\raggedright\begin{thebibliography}{100}

\bibitem{Georgi:1974sy}
H.~Georgi and S.~L. Glashow, ``{Unity of All Elementary Particle Forces},''
  \href{http://dx.doi.org/10.1103/PhysRevLett.32.438}{{\em Phys. Rev. Lett.}
  {\bfseries 32} (1974) 438--441}.

\bibitem{Fritzsch:1974nn}
H.~Fritzsch and P.~Minkowski, ``{Unified Interactions of Leptons and
  Hadrons},'' \href{http://dx.doi.org/10.1016/0003-4916(75)90211-0}{{\em Annals
  Phys.} {\bfseries 93} (1975) 193--266}.

\bibitem{Kawamura:1999nj}
Y.~Kawamura, ``{Gauge symmetry breaking from extra space S**1 / Z(2)},''
  \href{http://dx.doi.org/10.1143/PTP.103.613}{{\em Prog. Theor. Phys.}
  {\bfseries 103} (2000) 613--619},
  \href{http://arxiv.org/abs/hep-ph/9902423}{{\ttfamily arXiv:hep-ph/9902423}}.

\bibitem{Barbieri:2000vh}
R.~Barbieri, L.~J. Hall, and Y.~Nomura, ``{A Constrained standard model from a
  compact extra dimension},''
  \href{http://dx.doi.org/10.1103/PhysRevD.63.105007}{{\em Phys. Rev. D}
  {\bfseries 63} (2001) 105007},
  \href{http://arxiv.org/abs/hep-ph/0011311}{{\ttfamily arXiv:hep-ph/0011311}}.

\bibitem{Kawamura:2000ev}
Y.~Kawamura, ``{Triplet doublet splitting, proton stability and extra
  dimension},'' \href{http://dx.doi.org/10.1143/PTP.105.999}{{\em Prog. Theor.
  Phys.} {\bfseries 105} (2001) 999--1006},
  \href{http://arxiv.org/abs/hep-ph/0012125}{{\ttfamily arXiv:hep-ph/0012125}}.

\bibitem{Altarelli:2001qj}
G.~Altarelli and F.~Feruglio, ``{SU(5) grand unification in extra dimensions
  and proton decay},''
  \href{http://dx.doi.org/10.1016/S0370-2693(01)00650-5}{{\em Phys. Lett. B}
  {\bfseries 511} (2001) 257--264},
  \href{http://arxiv.org/abs/hep-ph/0102301}{{\ttfamily arXiv:hep-ph/0102301}}.

\bibitem{Hall:2001pg}
L.~J. Hall and Y.~Nomura, ``{Gauge unification in higher dimensions},''
  \href{http://dx.doi.org/10.1103/PhysRevD.64.055003}{{\em Phys. Rev. D}
  {\bfseries 64} (2001) 055003},
  \href{http://arxiv.org/abs/hep-ph/0103125}{{\ttfamily arXiv:hep-ph/0103125}}.

\bibitem{Nomura:2001mf}
Y.~Nomura, D.~Tucker-Smith, and N.~Weiner, ``{GUT breaking on the brane},''
  \href{http://dx.doi.org/10.1016/S0550-3213(01)00388-1}{{\em Nucl. Phys. B}
  {\bfseries 613} (2001) 147--166},
  \href{http://arxiv.org/abs/hep-ph/0104041}{{\ttfamily arXiv:hep-ph/0104041}}.

\bibitem{Hebecker:2001wq}
A.~Hebecker and J.~March-Russell, ``{A Minimal S**1 / (Z(2) x Z-prime (2))
  orbifold GUT},'' \href{http://dx.doi.org/10.1016/S0550-3213(01)00374-1}{{\em
  Nucl. Phys. B} {\bfseries 613} (2001) 3--16},
  \href{http://arxiv.org/abs/hep-ph/0106166}{{\ttfamily arXiv:hep-ph/0106166}}.

\bibitem{Hebecker:2001jb}
A.~Hebecker and J.~March-Russell, ``{The structure of GUT breaking by
  orbifolding},'' \href{http://dx.doi.org/10.1016/S0550-3213(02)00016-0}{{\em
  Nucl. Phys. B} {\bfseries 625} (2002) 128--150},
  \href{http://arxiv.org/abs/hep-ph/0107039}{{\ttfamily arXiv:hep-ph/0107039}}.

\bibitem{Asaka:2001eh}
T.~Asaka, W.~Buchmuller, and L.~Covi, ``{Gauge unification in
  six-dimensions},''
  \href{http://dx.doi.org/10.1016/S0370-2693(01)01324-7}{{\em Phys. Lett. B}
  {\bfseries 523} (2001) 199--204},
  \href{http://arxiv.org/abs/hep-ph/0108021}{{\ttfamily arXiv:hep-ph/0108021}}.

\bibitem{Hall:2001xr}
L.~J. Hall, Y.~Nomura, T.~Okui, and D.~Tucker-Smith, ``{SO(10) unified theories
  in six-dimensions},''
  \href{http://dx.doi.org/10.1103/PhysRevD.65.035008}{{\em Phys. Rev. D}
  {\bfseries 65} (2002) 035008},
  \href{http://arxiv.org/abs/hep-ph/0108071}{{\ttfamily arXiv:hep-ph/0108071}}.

\bibitem{Dermisek:2001hp}
R.~Dermisek and A.~Mafi, ``{SO(10) grand unification in five-dimensions: Proton
  decay and the mu problem},''
  \href{http://dx.doi.org/10.1103/PhysRevD.65.055002}{{\em Phys. Rev. D}
  {\bfseries 65} (2002) 055002},
  \href{http://arxiv.org/abs/hep-ph/0108139}{{\ttfamily arXiv:hep-ph/0108139}}.

\bibitem{Hall:2001xb}
L.~J. Hall and Y.~Nomura, ``{Gauge coupling unification from unified theories
  in higher dimensions},''
  \href{http://dx.doi.org/10.1103/PhysRevD.65.125012}{{\em Phys. Rev. D}
  {\bfseries 65} (2002) 125012},
  \href{http://arxiv.org/abs/hep-ph/0111068}{{\ttfamily arXiv:hep-ph/0111068}}.

\bibitem{Hall:2002ci}
L.~J. Hall and Y.~Nomura, ``{A Complete theory of grand unification in
  five-dimensions},'' \href{http://dx.doi.org/10.1103/PhysRevD.66.075004}{{\em
  Phys. Rev. D} {\bfseries 66} (2002) 075004},
  \href{http://arxiv.org/abs/hep-ph/0205067}{{\ttfamily arXiv:hep-ph/0205067}}.

\bibitem{Kim:2002im}
H.~D. Kim and S.~Raby, ``{Unification in 5D SO(10)},''
  \href{http://dx.doi.org/10.1088/1126-6708/2003/01/056}{{\em JHEP} {\bfseries
  01} (2003) 056}, \href{http://arxiv.org/abs/hep-ph/0212348}{{\ttfamily
  arXiv:hep-ph/0212348}}.

\bibitem{Kim:2004vk}
H.~D. Kim, S.~Raby, and L.~Schradin, ``{Quark and lepton masses in 5-D
  SO(10)},'' \href{http://dx.doi.org/10.1088/1126-6708/2005/05/036}{{\em JHEP}
  {\bfseries 05} (2005) 036},
  \href{http://arxiv.org/abs/hep-ph/0411328}{{\ttfamily arXiv:hep-ph/0411328}}.

\bibitem{Alciati:2005ur}
M.~L. Alciati, F.~Feruglio, Y.~Lin, and A.~Varagnolo, ``{Proton lifetime from
  SU(5) unification in extra dimensions},''
  \href{http://dx.doi.org/10.1088/1126-6708/2005/03/054}{{\em JHEP} {\bfseries
  03} (2005) 054}, \href{http://arxiv.org/abs/hep-ph/0501086}{{\ttfamily
  arXiv:hep-ph/0501086}}.

\bibitem{Agrawal:2022lsp}
P.~Agrawal, M.~Nee, and M.~Reig, ``{Axion couplings in grand unified
  theories},'' \href{http://dx.doi.org/10.1007/JHEP10(2022)141}{{\em JHEP}
  {\bfseries 10} (2022) 141}, \href{http://arxiv.org/abs/2206.07053}{{\ttfamily
  arXiv:2206.07053 [hep-ph]}}.

\bibitem{Agrawal:2024ejr}
P.~Agrawal, M.~Nee, and M.~Reig, ``{Axion couplings in heterotic string
  theory},'' \href{http://dx.doi.org/10.1007/JHEP02(2025)188}{{\em JHEP}
  {\bfseries 02} (2025) 188}, \href{http://arxiv.org/abs/2410.03820}{{\ttfamily
  arXiv:2410.03820 [hep-ph]}}.

\bibitem{Reig:2025dqb}
M.~Reig and T.~Weigand, ``{Testing the Heterotic String with the Axion-Photon
  Coupling},'' \href{http://arxiv.org/abs/2509.08042}{{\ttfamily
  arXiv:2509.08042 [hep-th]}}.

\bibitem{Marsh:2018dlj}
D.~J.~E. Marsh, K.-C. Fong, E.~W. Lentz, L.~Smejkal, and M.~N. Ali, ``{Proposal
  to Detect Dark Matter using Axionic Topological Antiferromagnets},''
  \href{http://dx.doi.org/10.1103/PhysRevLett.123.121601}{{\em Phys. Rev.
  Lett.} {\bfseries 123} no.~12, (2019) 121601},
  \href{http://arxiv.org/abs/1807.08810}{{\ttfamily arXiv:1807.08810
  [hep-ph]}}.

\bibitem{Lawson:2019brd}
M.~Lawson, A.~J. Millar, M.~Pancaldi, E.~Vitagliano, and F.~Wilczek, ``{Tunable
  axion plasma haloscopes},''
  \href{http://dx.doi.org/10.1103/PhysRevLett.123.141802}{{\em Phys. Rev.
  Lett.} {\bfseries 123} no.~14, (2019) 141802},
  \href{http://arxiv.org/abs/1904.11872}{{\ttfamily arXiv:1904.11872
  [hep-ph]}}.

\bibitem{Beurthey:2020yuq}
S.~Beurthey {\em et~al.}, ``{MADMAX Status Report},''
  \href{http://arxiv.org/abs/2003.10894}{{\ttfamily arXiv:2003.10894
  [physics.ins-det]}}.

\bibitem{Schutte-Engel:2021bqm}
J.~Sch{\"u}tte-Engel, D.~J.~E. Marsh, A.~J. Millar, A.~Sekine, F.~Chadha-Day,
  S.~Hoof, M.~N. Ali, K.-C. Fong, E.~Hardy, and L.~{\v{S}}mejkal, ``{Axion
  quasiparticles for axion dark matter detection},''
  \href{http://dx.doi.org/10.1088/1475-7516/2021/08/066}{{\em JCAP} {\bfseries
  08} (2021) 066}, \href{http://arxiv.org/abs/2102.05366}{{\ttfamily
  arXiv:2102.05366 [hep-ph]}}.

\bibitem{DMRadio:2022pkf}
{\bfseries DMRadio} Collaboration, L.~Brouwer {\em et~al.}, ``{Projected
  sensitivity of DMRadio-m3: A search for the QCD axion below
  1{\,}{\,}{\ensuremath{\mu}}eV},''
  \href{http://dx.doi.org/10.1103/PhysRevD.106.103008}{{\em Phys. Rev. D}
  {\bfseries 106} no.~10, (2022) 103008},
  \href{http://arxiv.org/abs/2204.13781}{{\ttfamily arXiv:2204.13781
  [hep-ex]}}.

\bibitem{Aja:2022csb}
B.~Aja {\em et~al.}, ``{The Canfranc Axion Detection Experiment (CADEx): search
  for axions at 90 GHz with Kinetic Inductance Detectors},''
  \href{http://dx.doi.org/10.1088/1475-7516/2022/11/044}{{\em JCAP} {\bfseries
  11} (2022) 044}, \href{http://arxiv.org/abs/2206.02980}{{\ttfamily
  arXiv:2206.02980 [hep-ex]}}.

\bibitem{Bourhill:2022alm}
J.~F. Bourhill, E.~C.~I. Paterson, M.~Goryachev, and M.~E. Tobar, ``{Searching
  for ultralight axions with twisted cavity resonators of anyon rotational
  symmetry with bulk modes of nonzero helicity},''
  \href{http://dx.doi.org/10.1103/PhysRevD.108.052014}{{\em Phys. Rev. D}
  {\bfseries 108} no.~5, (2023) 052014},
  \href{http://arxiv.org/abs/2208.01640}{{\ttfamily arXiv:2208.01640
  [hep-ph]}}.

\bibitem{ALPHA:2022rxj}
{\bfseries ALPHA} Collaboration, A.~J. Millar {\em et~al.}, ``{Searching for
  dark matter with plasma haloscopes},''
  \href{http://dx.doi.org/10.1103/PhysRevD.107.055013}{{\em Phys. Rev. D}
  {\bfseries 107} no.~5, (2023) 055013},
  \href{http://arxiv.org/abs/2210.00017}{{\ttfamily arXiv:2210.00017
  [hep-ph]}}.

\bibitem{Oshima:2023csb}
Y.~Oshima, H.~Fujimoto, J.~Kume, S.~Morisaki, K.~Nagano, T.~Fujita, I.~Obata,
  A.~Nishizawa, Y.~Michimura, and M.~Ando, ``{First results of axion dark
  matter search with DANCE},''
  \href{http://dx.doi.org/10.1103/PhysRevD.108.072005}{{\em Phys. Rev. D}
  {\bfseries 108} no.~7, (2023) 072005},
  \href{http://arxiv.org/abs/2303.03594}{{\ttfamily arXiv:2303.03594
  [hep-ex]}}.

\bibitem{DeMiguel:2023nmz}
{\bfseries DALI} Collaboration, J.~De~Miguel, J.~F. Hern{\'a}ndez-Cabrera,
  E.~Hern{\'a}ndez-Su{\'a}rez, E.~Joven-{\'A}lvarez, C.~Otani, and J.~A.
  Rubi{\~n}o-Mart{\'\i}n, ``{Discovery prospects with the Dark-photons {\&}
  Axion-like particles Interferometer},''
  \href{http://dx.doi.org/10.1103/PhysRevD.109.062002}{{\em Phys. Rev. D}
  {\bfseries 109} no.~6, (2024) 062002},
  \href{http://arxiv.org/abs/2303.03997}{{\ttfamily arXiv:2303.03997
  [hep-ph]}}.

\bibitem{Ahyoune:2023gfw}
S.~Ahyoune {\em et~al.}, ``{A Proposal for a Low-Frequency Axion Search in the
  1{\textendash}2 {\ensuremath{\mu}} eV Range and Below with the BabyIAXO
  Magnet},'' \href{http://dx.doi.org/10.1002/andp.202300326}{{\em Annalen
  Phys.} {\bfseries 535} no.~12, (2023) 2300326},
  \href{http://arxiv.org/abs/2306.17243}{{\ttfamily arXiv:2306.17243
  [physics.ins-det]}}.

\bibitem{Alesini:2023qed}
D.~Alesini {\em et~al.}, ``{The future search for low-frequency axions and new
  physics with the FLASH resonant cavity experiment at Frascati National
  Laboratories},'' \href{http://dx.doi.org/10.1016/j.dark.2023.101370}{{\em
  Phys. Dark Univ.} {\bfseries 42} (2023) 101370},
  \href{http://arxiv.org/abs/2309.00351}{{\ttfamily arXiv:2309.00351
  [physics.ins-det]}}.

\bibitem{BREAD:2023xhc}
{\bfseries BREAD} Collaboration, S.~Knirck {\em et~al.}, ``{First Results from
  a Broadband Search for Dark Photon Dark Matter in the 44 to
  52{\,}{\,}{\ensuremath{\mu}}eV Range with a Coaxial Dish Antenna},''
  \href{http://dx.doi.org/10.1103/PhysRevLett.132.131004}{{\em Phys. Rev.
  Lett.} {\bfseries 132} no.~13, (2024) 131004},
  \href{http://arxiv.org/abs/2310.13891}{{\ttfamily arXiv:2310.13891
  [hep-ex]}}.

\bibitem{CAST:2024eil}
{\bfseries CAST} Collaboration, K.~Altenm{\"u}ller {\em et~al.}, ``{New Upper
  Limit on the Axion-Photon Coupling with an Extended CAST Run with a Xe-Based
  Micromegas Detector},''
  \href{http://dx.doi.org/10.1103/PhysRevLett.133.221005}{{\em Phys. Rev.
  Lett.} {\bfseries 133} no.~22, (2024) 221005},
  \href{http://arxiv.org/abs/2406.16840}{{\ttfamily arXiv:2406.16840
  [hep-ex]}}.

\bibitem{Kalia:2024eml}
S.~Kalia, D.~Budker, D.~F.~J. Kimball, W.~Ji, Z.~Liu, A.~O. Sushkov,
  C.~Timberlake, H.~Ulbricht, A.~Vinante, and T.~Wang, ``{Ultralight dark
  matter detection with levitated ferromagnets},''
  \href{http://dx.doi.org/10.1103/PhysRevD.110.115029}{{\em Phys. Rev. D}
  {\bfseries 110} no.~11, (2024) 115029},
  \href{http://arxiv.org/abs/2408.15330}{{\ttfamily arXiv:2408.15330
  [hep-ph]}}.

\bibitem{Friel:2024shg}
M.~Friel, J.~W. Gjerloev, S.~Kalia, and A.~Zamora, ``{Search for ultralight
  dark matter in the SuperMAG high-fidelity dataset},''
  \href{http://dx.doi.org/10.1103/PhysRevD.110.115036}{{\em Phys. Rev. D}
  {\bfseries 110} no.~11, (2024) 115036},
  \href{http://arxiv.org/abs/2408.16045}{{\ttfamily arXiv:2408.16045
  [hep-ph]}}.

\bibitem{Minami:2020odp}
Y.~Minami and E.~Komatsu, ``{New Extraction of the Cosmic Birefringence from
  the Planck 2018 Polarization Data},''
  \href{http://dx.doi.org/10.1103/PhysRevLett.125.221301}{{\em Phys. Rev.
  Lett.} {\bfseries 125} no.~22, (2020) 221301},
  \href{http://arxiv.org/abs/2011.11254}{{\ttfamily arXiv:2011.11254
  [astro-ph.CO]}}.

\bibitem{Eskilt:2022cff}
J.~R. Eskilt and E.~Komatsu, ``{Improved constraints on cosmic birefringence
  from the WMAP and Planck cosmic microwave background polarization data},''
  \href{http://dx.doi.org/10.1103/PhysRevD.106.063503}{{\em Phys. Rev. D}
  {\bfseries 106} no.~6, (2022) 063503},
  \href{http://arxiv.org/abs/2205.13962}{{\ttfamily arXiv:2205.13962
  [astro-ph.CO]}}.

\bibitem{Diego-Palazuelos:2023mpy}
P.~Diego-Palazuelos, ``{Search for ultra-light axions with CMB polarization},''
\newblock 4, 2023.
\newblock \href{http://arxiv.org/abs/2304.03647}{{\ttfamily arXiv:2304.03647
  [astro-ph.CO]}}.

\bibitem{Diego-Palazuelos:2025dmh}
P.~Diego-Palazuelos and E.~Komatsu, ``{Cosmic Birefringence from the Atacama
  Cosmology Telescope Data Release 6},''
  \href{http://arxiv.org/abs/2509.13654}{{\ttfamily arXiv:2509.13654
  [astro-ph.CO]}}.

\bibitem{Font:1990uw}
A.~Font, L.~E. Ibanez, and F.~Quevedo, ``{Higher Level {Kac-Moody} String
  Models and Their Phenomenological Implications},''
  \href{http://dx.doi.org/10.1016/0550-3213(90)90393-R}{{\em Nucl. Phys. B}
  {\bfseries 345} (1990) 389--430}.

\bibitem{Dienes:1996yh}
K.~R. Dienes and J.~March-Russell, ``{Realizing higher level gauge symmetries
  in string theory: New embeddings for string GUTs},''
  \href{http://dx.doi.org/10.1016/0550-3213(96)00406-3}{{\em Nucl. Phys. B}
  {\bfseries 479} (1996) 113--172},
  \href{http://arxiv.org/abs/hep-th/9604112}{{\ttfamily arXiv:hep-th/9604112}}.

\bibitem{Blumenhagen:2009qh}
R.~Blumenhagen, M.~Cvetic, S.~Kachru, and T.~Weigand, ``{D-Brane Instantons in
  Type II Orientifolds},''
  \href{http://dx.doi.org/10.1146/annurev.nucl.010909.083113}{{\em Ann. Rev.
  Nucl. Part. Sci.} {\bfseries 59} (2009) 269--296},
  \href{http://arxiv.org/abs/0902.3251}{{\ttfamily arXiv:0902.3251 [hep-th]}}.

\bibitem{Polchinski:1998rr}
J.~Polchinski, \href{http://dx.doi.org/10.1017/CBO9780511618123}{{\em {String
  theory. Vol. 2: Superstring theory and beyond}}}.
\newblock Cambridge Monographs on Mathematical Physics. Cambridge University
  Press, 12, 2007.

\bibitem{Ibanez:2012zz}
L.~E. Ibanez and A.~M. Uranga, {\em {String theory and particle physics: An
  introduction to string phenomenology}}.
\newblock Cambridge University Press, 2, 2012.

\bibitem{Fan:2021ntg}
J.~Fan, K.~Fraser, M.~Reece, and J.~Stout, ``{Axion Mass from Magnetic Monopole
  Loops},'' \href{http://dx.doi.org/10.1103/PhysRevLett.127.131602}{{\em Phys.
  Rev. Lett.} {\bfseries 127} no.~13, (2021) 131602},
  \href{http://arxiv.org/abs/2105.09950}{{\ttfamily arXiv:2105.09950
  [hep-ph]}}.

\bibitem{GarciaGarcia:2025uub}
I.~Garcia~Garcia, M.~Kongsore, and K.~Van~Tilburg, ``{Dyon Loops and Abelian
  Instantons},'' \href{http://arxiv.org/abs/2506.14867}{{\ttfamily
  arXiv:2506.14867 [hep-th]}}.

\bibitem{Loladze:2025uvf}
V.~Loladze, A.~Platschorre, and M.~Reig, ``{Higher axion strings},''
  \href{http://dx.doi.org/10.1007/JHEP08(2025)182}{{\em JHEP} {\bfseries 08}
  (2025) 182}, \href{http://arxiv.org/abs/2503.18707}{{\ttfamily
  arXiv:2503.18707 [hep-ph]}}.

\bibitem{Petrossian-Byrne:2025mto}
R.~Petrossian-Byrne and G.~Villadoro, ``{Open string axiverse},''
  \href{http://dx.doi.org/10.1007/JHEP07(2025)049}{{\em JHEP} {\bfseries 07}
  (2025) 049}, \href{http://arxiv.org/abs/2503.16387}{{\ttfamily
  arXiv:2503.16387 [hep-ph]}}.

\bibitem{Uranga:933469}
A.~M. Uranga, ``{TASI lectures on string compactification, model building and
  fluxes},'' tech. rep., CERN, Geneva, 2005.
\newblock \url{https://cds.cern.ch/record/933469}.

\bibitem{Cicoli:2013cha}
M.~Cicoli, D.~Klevers, S.~Krippendorf, C.~Mayrhofer, F.~Quevedo, and
  R.~Valandro, ``{Explicit de Sitter Flux Vacua for Global String Models with
  Chiral Matter},'' \href{http://dx.doi.org/10.1007/JHEP05(2014)001}{{\em JHEP}
  {\bfseries 05} (2014) 001}, \href{http://arxiv.org/abs/1312.0014}{{\ttfamily
  arXiv:1312.0014 [hep-th]}}.

\bibitem{Cicoli:2013ana}
M.~Cicoli,
  \href{http://dx.doi.org/10.3204/DESY-PROC-2013-04/cicoli_michele}{``{Axion-like
  Particles from String Compactifications},''} in {\em {9th Patras Workshop on
  Axions, WIMPs and WISPs}}, pp.~235--242.
\newblock 2013.
\newblock \href{http://arxiv.org/abs/1309.6988}{{\ttfamily arXiv:1309.6988
  [hep-th]}}.

\bibitem{Allahverdi:2014ppa}
R.~Allahverdi, M.~Cicoli, B.~Dutta, and K.~Sinha, ``{Correlation between Dark
  Matter and Dark Radiation in String Compactifications},''
  \href{http://dx.doi.org/10.1088/1475-7516/2014/10/002}{{\em JCAP} {\bfseries
  10} (2014) 002}, \href{http://arxiv.org/abs/1401.4364}{{\ttfamily
  arXiv:1401.4364 [hep-ph]}}.

\bibitem{Dine:1987xk}
M.~Dine, N.~Seiberg, and E.~Witten, ``{Fayet-Iliopoulos Terms in String
  Theory},'' \href{http://dx.doi.org/10.1016/0550-3213(87)90395-6}{{\em Nucl.
  Phys. B} {\bfseries 289} (1987) 589--598}.

\bibitem{Choi:2003wr}
K.-w. Choi, ``{A QCD axion from higher dimensional gauge field},''
  \href{http://dx.doi.org/10.1103/PhysRevLett.92.101602}{{\em Phys. Rev. Lett.}
  {\bfseries 92} (2004) 101602},
  \href{http://arxiv.org/abs/hep-ph/0308024}{{\ttfamily arXiv:hep-ph/0308024}}.

\bibitem{Arkani-Hamed:1999ylh}
N.~Arkani-Hamed and M.~Schmaltz, ``{Hierarchies without symmetries from extra
  dimensions},'' \href{http://dx.doi.org/10.1103/PhysRevD.61.033005}{{\em Phys.
  Rev. D} {\bfseries 61} (2000) 033005},
  \href{http://arxiv.org/abs/hep-ph/9903417}{{\ttfamily arXiv:hep-ph/9903417}}.

\bibitem{Aldazabal:2000cn}
G.~Aldazabal, S.~Franco, L.~E. Ibanez, R.~Rabadan, and A.~M. Uranga,
  ``{Intersecting brane worlds},''
  \href{http://dx.doi.org/10.1088/1126-6708/2001/02/047}{{\em JHEP} {\bfseries
  02} (2001) 047}, \href{http://arxiv.org/abs/hep-ph/0011132}{{\ttfamily
  arXiv:hep-ph/0011132}}.

\bibitem{Gherghetta:2000qt}
T.~Gherghetta and A.~Pomarol, ``{Bulk fields and supersymmetry in a slice of
  AdS},'' \href{http://dx.doi.org/10.1016/S0550-3213(00)00392-8}{{\em Nucl.
  Phys. B} {\bfseries 586} (2000) 141--162},
  \href{http://arxiv.org/abs/hep-ph/0003129}{{\ttfamily arXiv:hep-ph/0003129}}.

\bibitem{Kahn:2016aff}
Y.~Kahn, B.~R. Safdi, and J.~Thaler, ``{Broadband and Resonant Approaches to
  Axion Dark Matter Detection},''
  \href{http://dx.doi.org/10.1103/PhysRevLett.117.141801}{{\em Phys. Rev.
  Lett.} {\bfseries 117} no.~14, (2016) 141801},
  \href{http://arxiv.org/abs/1602.01086}{{\ttfamily arXiv:1602.01086
  [hep-ph]}}.

\bibitem{Ouellet:2018beu}
J.~L. Ouellet {\em et~al.}, ``{First Results from ABRACADABRA-10 cm: A Search
  for Sub-$\mu$eV Axion Dark Matter},''
  \href{http://dx.doi.org/10.1103/PhysRevLett.122.121802}{{\em Phys. Rev.
  Lett.} {\bfseries 122} no.~12, (2019) 121802},
  \href{http://arxiv.org/abs/1810.12257}{{\ttfamily arXiv:1810.12257
  [hep-ex]}}.

\bibitem{Ouellet:2019tlz}
J.~L. Ouellet {\em et~al.}, ``{Design and implementation of the ABRACADABRA-10
  cm axion dark matter search},''
  \href{http://dx.doi.org/10.1103/PhysRevD.99.052012}{{\em Phys. Rev. D}
  {\bfseries 99} no.~5, (2019) 052012},
  \href{http://arxiv.org/abs/1901.10652}{{\ttfamily arXiv:1901.10652
  [physics.ins-det]}}.

\bibitem{Salemi:2021gck}
C.~P. Salemi {\em et~al.}, ``{Search for Low-Mass Axion Dark Matter with
  ABRACADABRA-10~cm},''
  \href{http://dx.doi.org/10.1103/PhysRevLett.127.081801}{{\em Phys. Rev.
  Lett.} {\bfseries 127} no.~8, (2021) 081801},
  \href{http://arxiv.org/abs/2102.06722}{{\ttfamily arXiv:2102.06722
  [hep-ex]}}.

\bibitem{DMRadio:2022jfv}
{\bfseries DMRadio} Collaboration, L.~Brouwer {\em et~al.}, ``{Proposal for a
  definitive search for GUT-scale QCD axions},''
  \href{http://dx.doi.org/10.1103/PhysRevD.106.112003}{{\em Phys. Rev. D}
  {\bfseries 106} no.~11, (2022) 112003},
  \href{http://arxiv.org/abs/2203.11246}{{\ttfamily arXiv:2203.11246
  [hep-ex]}}.

\bibitem{Benabou:2022qpv}
J.~N. Benabou, J.~W. Foster, Y.~Kahn, B.~R. Safdi, and C.~P. Salemi,
  ``{Lumped-element axion dark matter detection beyond the magnetoquasistatic
  limit},'' \href{http://dx.doi.org/10.1103/PhysRevD.108.035009}{{\em Phys.
  Rev. D} {\bfseries 108} no.~3, (2023) 035009},
  \href{http://arxiv.org/abs/2211.00008}{{\ttfamily arXiv:2211.00008
  [hep-ph]}}.

\bibitem{DMRadio:2023igr}
{\bfseries DMRadio} Collaboration, A.~AlShirawi {\em et~al.},
  ``{Electromagnetic modeling and science reach of DMRadio-m3},''
  \href{http://dx.doi.org/10.1103/bwgd-kxsb}{{\em Phys. Rev. D} {\bfseries 112}
  no.~5, (2025) 052001}, \href{http://arxiv.org/abs/2302.14084}{{\ttfamily
  arXiv:2302.14084 [hep-ex]}}.

\bibitem{Berlin:2019ahk}
A.~Berlin, R.~T. D'Agnolo, S.~A.~R. Ellis, C.~Nantista, J.~Neilson,
  P.~Schuster, S.~Tantawi, N.~Toro, and K.~Zhou, ``{Axion Dark Matter Detection
  by Superconducting Resonant Frequency Conversion},''
  \href{http://dx.doi.org/10.1007/JHEP07(2020)088}{{\em JHEP} {\bfseries 07}
  no.~07, (2020) 088}, \href{http://arxiv.org/abs/1912.11048}{{\ttfamily
  arXiv:1912.11048 [hep-ph]}}.

\bibitem{Giaccone:2022pke}
B.~Giaccone {\em et~al.}, ``{Design of axion and axion dark matter searches
  based on ultra high Q SRF cavities},''
  \href{http://arxiv.org/abs/2207.11346}{{\ttfamily arXiv:2207.11346
  [hep-ex]}}.

\bibitem{Graham:2013gfa}
P.~W. Graham and S.~Rajendran, ``{New Observables for Direct Detection of Axion
  Dark Matter},'' \href{http://dx.doi.org/10.1103/PhysRevD.88.035023}{{\em
  Phys. Rev. D} {\bfseries 88} (2013) 035023},
  \href{http://arxiv.org/abs/1306.6088}{{\ttfamily arXiv:1306.6088 [hep-ph]}}.

\bibitem{Budker:2013hfa}
D.~Budker, P.~W. Graham, M.~Ledbetter, S.~Rajendran, and A.~Sushkov,
  ``{Proposal for a Cosmic Axion Spin Precession Experiment (CASPEr)},''
  \href{http://dx.doi.org/10.1103/PhysRevX.4.021030}{{\em Phys. Rev. X}
  {\bfseries 4} no.~2, (2014) 021030},
  \href{http://arxiv.org/abs/1306.6089}{{\ttfamily arXiv:1306.6089 [hep-ph]}}.

\bibitem{JacksonKimball:2017elr}
D.~F. Jackson~Kimball {\em et~al.}, ``{Overview of the Cosmic Axion Spin
  Precession Experiment (CASPEr)},''
  \href{http://dx.doi.org/10.1007/978-3-030-43761-9_13}{{\em Springer Proc.
  Phys.} {\bfseries 245} (2020) 105--121},
  \href{http://arxiv.org/abs/1711.08999}{{\ttfamily arXiv:1711.08999
  [physics.ins-det]}}.

\bibitem{Aybas:2021cdk}
D.~Aybas, H.~Bekker, J.~W. Blanchard, D.~Budker, G.~P. Centers, N.~L. Figueroa,
  A.~V. Gramolin, D.~F.~J. Kimball, A.~Wickenbrock, and A.~O. Sushkov,
  ``{Quantum sensitivity limits of nuclear magnetic resonance experiments
  searching for new fundamental physics},''
  \href{http://dx.doi.org/10.1088/2058-9565/abfbbc}{{\em Quantum Sci. Technol.}
  {\bfseries 6} no.~3, (2021) 034007},
  \href{http://arxiv.org/abs/2103.06284}{{\ttfamily arXiv:2103.06284
  [quant-ph]}}.

\bibitem{Dror:2022xpi}
J.~A. Dror, S.~Gori, J.~M. Leedom, and N.~L. Rodd, ``{Sensitivity of
  Spin-Precession Axion Experiments},''
  \href{http://dx.doi.org/10.1103/PhysRevLett.130.181801}{{\em Phys. Rev.
  Lett.} {\bfseries 130} no.~18, (2023) 181801},
  \href{http://arxiv.org/abs/2210.06481}{{\ttfamily arXiv:2210.06481
  [hep-ph]}}.

\bibitem{Witten:1985xc}
E.~Witten, ``{Symmetry breaking patterns in superstring models},''
  \href{http://dx.doi.org/10.1016/0550-3213(85)90603-0}{{\em Nucl. Phys. B}
  {\bfseries 258} (1985) 75--100}.

\bibitem{Green:1987mn}
M.~B. Green, J.~H. Schwarz, and E.~Witten, {\em {SUPERSTRING THEORY. VOL. 2:
  LOOP AMPLITUDES, ANOMALIES AND PHENOMENOLOGY}}.
\newblock 7, 1988.

\bibitem{Barr:1981qv}
S.~M. Barr, ``{A New Symmetry Breaking Pattern for SO(10) and Proton Decay},''
  \href{http://dx.doi.org/10.1016/0370-2693(82)90966-2}{{\em Phys. Lett. B}
  {\bfseries 112} (1982) 219--222}.

\bibitem{Derendinger:1983aj}
J.~P. Derendinger, J.~E. Kim, and D.~V. Nanopoulos, ``{Anti-SU(5)},''
  \href{http://dx.doi.org/10.1016/0370-2693(84)91238-3}{{\em Phys. Lett. B}
  {\bfseries 139} (1984) 170--176}.

\bibitem{Huang:2016dtj}
J.~Huang and J.~March-Russell, ``{Unified Maximally Natural Supersymmetry},''
  \href{http://arxiv.org/abs/1607.08622}{{\ttfamily arXiv:1607.08622
  [hep-ph]}}.

\bibitem{Randall:1999ee}
L.~Randall and R.~Sundrum, ``{A Large mass hierarchy from a small extra
  dimension},'' \href{http://dx.doi.org/10.1103/PhysRevLett.83.3370}{{\em Phys.
  Rev. Lett.} {\bfseries 83} (1999) 3370--3373},
  \href{http://arxiv.org/abs/hep-ph/9905221}{{\ttfamily arXiv:hep-ph/9905221}}.

\bibitem{Randall:1999vf}
L.~Randall and R.~Sundrum, ``{An Alternative to compactification},''
  \href{http://dx.doi.org/10.1103/PhysRevLett.83.4690}{{\em Phys. Rev. Lett.}
  {\bfseries 83} (1999) 4690--4693},
  \href{http://arxiv.org/abs/hep-th/9906064}{{\ttfamily arXiv:hep-th/9906064}}.

\bibitem{Agrawal:2017ksf}
P.~Agrawal and K.~Howe, ``{Factoring the Strong CP Problem},''
  \href{http://dx.doi.org/10.1007/JHEP12(2018)029}{{\em JHEP} {\bfseries 12}
  (2018) 029}, \href{http://arxiv.org/abs/1710.04213}{{\ttfamily
  arXiv:1710.04213 [hep-ph]}}.

\bibitem{Leedom:2025mlr}
J.~M. Leedom, M.~Putti, and A.~Westphal, ``{Towards a Heterotic Axiverse},''
  \href{http://arxiv.org/abs/2509.03578}{{\ttfamily arXiv:2509.03578
  [hep-th]}}.

\bibitem{Arkani-Hamed:2003xts}
N.~Arkani-Hamed, H.-C. Cheng, P.~Creminelli, and L.~Randall, ``{Extra natural
  inflation},'' \href{http://dx.doi.org/10.1103/PhysRevLett.90.221302}{{\em
  Phys. Rev. Lett.} {\bfseries 90} (2003) 221302},
  \href{http://arxiv.org/abs/hep-th/0301218}{{\ttfamily arXiv:hep-th/0301218}}.

\bibitem{Craig:2024dnl}
N.~Craig and M.~Kongsore, ``{High-quality axions from higher-form symmetries in
  extra dimensions},''
  \href{http://dx.doi.org/10.1103/PhysRevD.111.015047}{{\em Phys. Rev. D}
  {\bfseries 111} no.~1, (2025) 015047},
  \href{http://arxiv.org/abs/2408.10295}{{\ttfamily arXiv:2408.10295
  [hep-ph]}}.

\bibitem{Gendler:2023kjt}
N.~Gendler, D.~J.~E. Marsh, L.~McAllister, and J.~Moritz, ``{Glimmers from the
  axiverse},'' \href{http://dx.doi.org/10.1088/1475-7516/2024/09/071}{{\em
  JCAP} {\bfseries 09} (2024) 071},
  \href{http://arxiv.org/abs/2309.13145}{{\ttfamily arXiv:2309.13145
  [hep-th]}}.

\bibitem{Yin:2025amn}
Z.~Yin, H.~Cheng, E.~Di~Valentino, N.~Gendler, D.~J.~E. Marsh, and
  L.~Visinelli, ``{Constraining the axiverse with reionization},''
  \href{http://arxiv.org/abs/2507.03535}{{\ttfamily arXiv:2507.03535
  [hep-ph]}}.

\bibitem{Benabou:2024jlj}
J.~N. Benabou, C.~A. Manzari, Y.~Park, G.~Prabhakar, B.~R. Safdi, and
  I.~Savoray, ``{Time-delayed gamma-ray signatures of heavy axions from
  core-collapse supernovae},''
  \href{http://dx.doi.org/10.1103/PhysRevD.111.095029}{{\em Phys. Rev. D}
  {\bfseries 111} no.~9, (2025) 095029},
  \href{http://arxiv.org/abs/2412.13247}{{\ttfamily arXiv:2412.13247
  [hep-ph]}}.

\bibitem{Jaeckel:2015jla}
J.~Jaeckel and M.~Spannowsky, ``{Probing MeV to 90 GeV axion-like particles
  with LEP and LHC},''
  \href{http://dx.doi.org/10.1016/j.physletb.2015.12.037}{{\em Phys. Lett. B}
  {\bfseries 753} (2016) 482--487},
  \href{http://arxiv.org/abs/1509.00476}{{\ttfamily arXiv:1509.00476
  [hep-ph]}}.

\bibitem{Bauer:2017ris}
M.~Bauer, M.~Neubert, and A.~Thamm, ``{Collider Probes of Axion-Like
  Particles},'' \href{http://dx.doi.org/10.1007/JHEP12(2017)044}{{\em JHEP}
  {\bfseries 12} (2017) 044}, \href{http://arxiv.org/abs/1708.00443}{{\ttfamily
  arXiv:1708.00443 [hep-ph]}}.

\bibitem{Bedi:2025hbz}
R.~Bedi, T.~Gherghetta, S.~Kumar, P.~Li, and Z.~Liu, ``{Heavy QCD Axions at
  High-Energy Muon Colliders},''
  \href{http://arxiv.org/abs/2509.10605}{{\ttfamily arXiv:2509.10605
  [hep-ph]}}.

\bibitem{Foster:2022ajl}
J.~W. Foster, S.~Kumar, B.~R. Safdi, and Y.~Soreq, ``{Dark Grand Unification in
  the axiverse: decaying axion dark matter and spontaneous baryogenesis},''
  \href{http://dx.doi.org/10.1007/JHEP12(2022)119}{{\em JHEP} {\bfseries 12}
  (2022) 119}, \href{http://arxiv.org/abs/2208.10504}{{\ttfamily
  arXiv:2208.10504 [hep-ph]}}.

\bibitem{Gavela:2023tzu}
B.~Gavela, P.~Qu{\'\i}lez, and M.~Ramos, ``{The QCD axion sum rule},''
  \href{http://dx.doi.org/10.1007/JHEP04(2024)056}{{\em JHEP} {\bfseries 04}
  (2024) 056}, \href{http://arxiv.org/abs/2305.15465}{{\ttfamily
  arXiv:2305.15465 [hep-ph]}}.

\bibitem{Dunsky:2025sgz}
D.~I. Dunsky, C.~A. Manzari, P.~Qu{\'\i}lez, M.~Ramos, and P.~S{\o}rensen,
  ``{Resonant Landau-Zener Conversion In Multi-Axion Systems},''
  \href{http://arxiv.org/abs/2507.06287}{{\ttfamily arXiv:2507.06287
  [hep-ph]}}.

\bibitem{tHooft:1976snw}
G.~'t~Hooft, ``{Computation of the Quantum Effects Due to a Four-Dimensional
  Pseudoparticle},'' \href{http://dx.doi.org/10.1103/PhysRevD.14.3432}{{\em
  Phys. Rev. D} {\bfseries 14} (1976) 3432--3450}. [Erratum: Phys.Rev.D 18,
  2199 (1978)].

\bibitem{Caputo:2024oqc}
A.~Caputo and G.~Raffelt, ``{Astrophysical Axion Bounds: The 2024 Edition},''
  \href{http://dx.doi.org/10.22323/1.454.0041}{{\em PoS} {\bfseries
  COSMICWISPers} (2024) 041}, \href{http://arxiv.org/abs/2401.13728}{{\ttfamily
  arXiv:2401.13728 [hep-ph]}}.

\bibitem{AxionLimits}
C.~O'Hare, ``cajohare/axionlimits: Axionlimits.''
  \url{https://cajohare.github.io/AxionLimits/}, July, 2020.

\bibitem{Choi:2011xt}
K.~Choi, K.~S. Jeong, K.-I. Okumura, and M.~Yamaguchi, ``{Mixed Mediation of
  Supersymmetry Breaking with Anomalous U(1) Gauge Symmetry},''
  \href{http://dx.doi.org/10.1007/JHEP06(2011)049}{{\em JHEP} {\bfseries 06}
  (2011) 049}, \href{http://arxiv.org/abs/1104.3274}{{\ttfamily arXiv:1104.3274
  [hep-ph]}}.

\bibitem{Choi:2014uaa}
K.~Choi, K.~S. Jeong, and M.-S. Seo, ``{String theoretic QCD axions in the
  light of PLANCK and BICEP2},''
  \href{http://dx.doi.org/10.1007/JHEP07(2014)092}{{\em JHEP} {\bfseries 07}
  (2014) 092}, \href{http://arxiv.org/abs/1404.3880}{{\ttfamily arXiv:1404.3880
  [hep-th]}}.

\bibitem{Buchbinder:2014qca}
E.~I. Buchbinder, A.~Constantin, and A.~Lukas, ``{Heterotic QCD axion},''
  \href{http://dx.doi.org/10.1103/PhysRevD.91.046010}{{\em Phys. Rev. D}
  {\bfseries 91} no.~4, (2015) 046010},
  \href{http://arxiv.org/abs/1412.8696}{{\ttfamily arXiv:1412.8696 [hep-th]}}.

\bibitem{Green:1984sg}
M.~B. Green and J.~H. Schwarz, ``{Anomaly Cancellation in Supersymmetric D=10
  Gauge Theory and Superstring Theory},''
  \href{http://dx.doi.org/10.1016/0370-2693(84)91565-X}{{\em Phys. Lett. B}
  {\bfseries 149} (1984) 117--122}.

\bibitem{Benabou:2023npn}
J.~N. Benabou, Q.~Bonnefoy, M.~Buschmann, S.~Kumar, and B.~R. Safdi,
  ``{Cosmological dynamics of string theory axion strings},''
  \href{http://dx.doi.org/10.1103/PhysRevD.110.035021}{{\em Phys. Rev. D}
  {\bfseries 110} no.~3, (2024) 035021},
  \href{http://arxiv.org/abs/2312.08425}{{\ttfamily arXiv:2312.08425
  [hep-ph]}}.

\bibitem{Saikawa:2024bta}
K.~Saikawa, J.~Redondo, A.~Vaquero, and M.~Kaltschmidt, ``{Spectrum of global
  string networks and the axion dark matter mass},''
  \href{http://dx.doi.org/10.1088/1475-7516/2024/10/043}{{\em JCAP} {\bfseries
  10} (2024) 043}, \href{http://arxiv.org/abs/2401.17253}{{\ttfamily
  arXiv:2401.17253 [hep-ph]}}.

\bibitem{Correia:2024cpk}
J.~Correia, M.~Hindmarsh, J.~Lizarraga, A.~Lopez-Eiguren, K.~Rummukainen, and
  J.~Urrestilla, ``{Scaling density of axion strings in terasite
  simulations},'' \href{http://dx.doi.org/10.1103/PhysRevD.111.063532}{{\em
  Phys. Rev. D} {\bfseries 111} no.~6, (2025) 063532},
  \href{http://arxiv.org/abs/2410.18064}{{\ttfamily arXiv:2410.18064
  [hep-ph]}}.

\bibitem{Benabou:2024msj}
J.~N. Benabou, M.~Buschmann, J.~W. Foster, and B.~R. Safdi, ``{Axion Mass
  Prediction from Adaptive Mesh Refinement Cosmological Lattice Simulations},''
  \href{http://dx.doi.org/10.1103/6v21-d6sj}{{\em Phys. Rev. Lett.} {\bfseries
  134} no.~24, (2025) 241003},
  \href{http://arxiv.org/abs/2412.08699}{{\ttfamily arXiv:2412.08699
  [hep-ph]}}.

\bibitem{Hill:2000mu}
C.~T. Hill, S.~Pokorski, and J.~Wang, ``{Gauge Invariant Effective Lagrangian
  for Kaluza-Klein Modes},''
  \href{http://dx.doi.org/10.1103/PhysRevD.64.105005}{{\em Phys. Rev. D}
  {\bfseries 64} (2001) 105005},
  \href{http://arxiv.org/abs/hep-th/0104035}{{\ttfamily arXiv:hep-th/0104035}}.

\bibitem{Arkani-Hamed:2001kyx}
N.~Arkani-Hamed, A.~G. Cohen, and H.~Georgi, ``{(De)constructing dimensions},''
  \href{http://dx.doi.org/10.1103/PhysRevLett.86.4757}{{\em Phys. Rev. Lett.}
  {\bfseries 86} (2001) 4757--4761},
  \href{http://arxiv.org/abs/hep-th/0104005}{{\ttfamily arXiv:hep-th/0104005}}.

\bibitem{Csaki:2001qm}
C.~Csaki, G.~D. Kribs, and J.~Terning, ``{4-D models of Scherk-Schwarz GUT
  breaking via deconstruction},''
  \href{http://dx.doi.org/10.1103/PhysRevD.65.015004}{{\em Phys. Rev. D}
  {\bfseries 65} (2002) 015004},
  \href{http://arxiv.org/abs/hep-ph/0107266}{{\ttfamily arXiv:hep-ph/0107266}}.

\bibitem{Poppitz:2002ac}
E.~Poppitz and Y.~Shirman, ``{The Strength of small instanton amplitudes in
  gauge theories with compact extra dimensions},''
  \href{http://dx.doi.org/10.1088/1126-6708/2002/07/041}{{\em JHEP} {\bfseries
  07} (2002) 041}, \href{http://arxiv.org/abs/hep-th/0204075}{{\ttfamily
  arXiv:hep-th/0204075}}.

\bibitem{Conlon:2006tq}
J.~P. Conlon, ``{The QCD axion and moduli stabilisation},''
  \href{http://dx.doi.org/10.1088/1126-6708/2006/05/078}{{\em JHEP} {\bfseries
  05} (2006) 078}, \href{http://arxiv.org/abs/hep-th/0602233}{{\ttfamily
  arXiv:hep-th/0602233}}.

\bibitem{Svrcek:2006yi}
P.~Svrcek and E.~Witten, ``{Axions In String Theory},''
  \href{http://dx.doi.org/10.1088/1126-6708/2006/06/051}{{\em JHEP} {\bfseries
  06} (2006) 051}, \href{http://arxiv.org/abs/hep-th/0605206}{{\ttfamily
  arXiv:hep-th/0605206}}.

\bibitem{Arvanitaki:2009fg}
A.~Arvanitaki, S.~Dimopoulos, S.~Dubovsky, N.~Kaloper, and J.~March-Russell,
  ``{String Axiverse},''
  \href{http://dx.doi.org/10.1103/PhysRevD.81.123530}{{\em Phys. Rev. D}
  {\bfseries 81} (2010) 123530},
  \href{http://arxiv.org/abs/0905.4720}{{\ttfamily arXiv:0905.4720 [hep-th]}}.

\bibitem{Cicoli:2012sz}
M.~Cicoli, M.~Goodsell, and A.~Ringwald, ``{The type IIB string axiverse and
  its low-energy phenomenology},''
  \href{http://dx.doi.org/10.1007/JHEP10(2012)146}{{\em JHEP} {\bfseries 10}
  (2012) 146}, \href{http://arxiv.org/abs/1206.0819}{{\ttfamily arXiv:1206.0819
  [hep-th]}}.

\bibitem{Hebecker:2018yxs}
A.~Hebecker, S.~Leonhardt, J.~Moritz, and A.~Westphal, ``{Thraxions: Ultralight
  Throat Axions},'' \href{http://dx.doi.org/10.1007/JHEP04(2019)158}{{\em JHEP}
  {\bfseries 04} (2019) 158}, \href{http://arxiv.org/abs/1812.03999}{{\ttfamily
  arXiv:1812.03999 [hep-th]}}.

\bibitem{Demirtas:2018akl}
M.~Demirtas, C.~Long, L.~McAllister, and M.~Stillman, ``{The Kreuzer-Skarke
  Axiverse},'' \href{http://dx.doi.org/10.1007/JHEP04(2020)138}{{\em JHEP}
  {\bfseries 04} (2020) 138}, \href{http://arxiv.org/abs/1808.01282}{{\ttfamily
  arXiv:1808.01282 [hep-th]}}.

\bibitem{Halverson:2019cmy}
J.~Halverson, C.~Long, B.~Nelson, and G.~Salinas, ``{Towards string theory
  expectations for photon couplings to axionlike particles},''
  \href{http://dx.doi.org/10.1103/PhysRevD.100.106010}{{\em Phys. Rev. D}
  {\bfseries 100} no.~10, (2019) 106010},
  \href{http://arxiv.org/abs/1909.05257}{{\ttfamily arXiv:1909.05257
  [hep-th]}}.

\bibitem{Mehta:2021pwf}
V.~M. Mehta, M.~Demirtas, C.~Long, D.~J.~E. Marsh, L.~McAllister, and M.~J.
  Stott, ``{Superradiance in string theory},''
  \href{http://dx.doi.org/10.1088/1475-7516/2021/07/033}{{\em JCAP} {\bfseries
  07} (2021) 033}, \href{http://arxiv.org/abs/2103.06812}{{\ttfamily
  arXiv:2103.06812 [hep-th]}}.

\bibitem{Demirtas:2021gsq}
M.~Demirtas, N.~Gendler, C.~Long, L.~McAllister, and J.~Moritz, ``{PQ
  axiverse},'' \href{http://dx.doi.org/10.1007/JHEP06(2023)092}{{\em JHEP}
  {\bfseries 06} (2023) 092}, \href{http://arxiv.org/abs/2112.04503}{{\ttfamily
  arXiv:2112.04503 [hep-th]}}.

\bibitem{Gendler:2023hwg}
N.~Gendler, O.~Janssen, M.~Kleban, J.~La~Madrid, and V.~M. Mehta, ``{Axion
  minima in string theory},''
  \href{http://dx.doi.org/10.1007/JHEP02(2025)134}{{\em JHEP} {\bfseries 02}
  (2025) 134}, \href{http://arxiv.org/abs/2309.01831}{{\ttfamily
  arXiv:2309.01831 [hep-th]}}.

\bibitem{Reece:2024wrn}
M.~Reece, ``{Extra-dimensional axion expectations},''
  \href{http://dx.doi.org/10.1007/JHEP07(2025)130}{{\em JHEP} {\bfseries 07}
  (2025) 130}, \href{http://arxiv.org/abs/2406.08543}{{\ttfamily
  arXiv:2406.08543 [hep-ph]}}.

\bibitem{Gendler:2024adn}
N.~Gendler and D.~J.~E. Marsh, ``{Possible Implications of QCD Axion Dark
  Matter Constraints from Helioscopes and Haloscopes for the String Theory
  Landscape},'' \href{http://dx.doi.org/10.1103/PhysRevLett.134.081602}{{\em
  Phys. Rev. Lett.} {\bfseries 134} no.~8, (2025) 081602},
  \href{http://arxiv.org/abs/2407.07143}{{\ttfamily arXiv:2407.07143
  [hep-th]}}.

\bibitem{Benabou:2025kgx}
J.~N. Benabou, K.~Fraser, M.~Reig, and B.~R. Safdi, ``{String theory and grand
  unification suggest a submicroelectronvolt QCD axion},''
  \href{http://dx.doi.org/10.1103/lthr-97lm}{{\em Phys. Rev. D} {\bfseries 112}
  no.~6, (2025) 066003}, \href{http://arxiv.org/abs/2505.15884}{{\ttfamily
  arXiv:2505.15884 [hep-ph]}}.

\bibitem{Cheng:2025ggf}
J.~Cheng and N.~Gendler, ``{Universality in the Axiverse},''
  \href{http://arxiv.org/abs/2507.12516}{{\ttfamily arXiv:2507.12516
  [hep-th]}}.

\bibitem{Fallon:2025lvn}
S.~V.~P. Fallon, J.~Halverson, L.~McAllister, and Y.~Zhu, ``{F-theory
  Axiverse},'' \href{http://arxiv.org/abs/2511.20458}{{\ttfamily
  arXiv:2511.20458 [hep-th]}}.

\bibitem{Horava:1995qa}
P.~Horava and E.~Witten, ``{Heterotic and Type I string dynamics from eleven
  dimensions},'' \href{http://dx.doi.org/10.1201/9781482268737-35}{{\em Nucl.
  Phys. B} {\bfseries 460} (1996) 506--524},
  \href{http://arxiv.org/abs/hep-th/9510209}{{\ttfamily arXiv:hep-th/9510209}}.

\bibitem{Horava:1996ma}
P.~Horava and E.~Witten, ``{Eleven-dimensional supergravity on a manifold with
  boundary},'' \href{http://dx.doi.org/10.1016/0550-3213(96)00308-2}{{\em Nucl.
  Phys. B} {\bfseries 475} (1996) 94--114},
  \href{http://arxiv.org/abs/hep-th/9603142}{{\ttfamily arXiv:hep-th/9603142}}.

\end{thebibliography}\endgroup

\end{document}